\documentclass[12pt]{article}
\usepackage{amsmath,amssymb,theorem,cite,epsfig,url,psfrag,eepic,amsmath,mathtools,mathrsfs,amsbsy,dsfont,esint,braket,cancel}
\usepackage{indentfirst}
\usepackage{graphicx}
\usepackage{tikz,pgfplots}
\usepackage{bm,upgreek}
\usepackage[bookmarks=true,hyperfigures=true,colorlinks=true,linkcolor=black,citecolor=black,urlcolor=black,bookmarksnumbered,hidelinks]{hyperref}

\def\XXint#1#2#3{{\setbox0=\hbox{$#1{#2#3}{\int}$}
     \vcenter{\hbox{$#2#3$}}\kern-.5\wd0}}

\advance \topmargin by -\headheight
\advance \topmargin by -\headsep     
\evensidemargin \oddsidemargin
\def\Maketitle{{\def\newpage{}\maketitle}}
\numberwithin{equation}{section}
\begin{document}
\title{\textbf{Meson mass spectrum in QCD$_2$ 't Hooft's model
}\vspace*{0.3cm}}
\date{}
\author{Alexey Litvinov$^{1,2}$ and Pavel Meshcheriakov$^{1,2,3}$
\\[\medskipamount]
\parbox[t]{0.85\textwidth}{\normalsize\it\centerline{1. Krichever Center, Skolkovo Institute of Science and Technology, 121205 Moscow, Russia}}
\\
\parbox[t]{0.85\textwidth}{\normalsize\it\centerline{2. Landau Institute for Theoretical Physics, 142432 Chernogolovka, Russia}}
\\
\parbox[t]{0.85\textwidth}{\normalsize\it\centerline{3. Moscow Institute of Physics and Technology, 141700 Dolgoprudny, Russia}}}
\Maketitle
\begin{abstract}
    We study the spectrum of meson masses in large $N_c$ QCD$_2$ governed by celebrated 't Hooft's integral equation. We generalize analytical methods proposed by Fateev, Lukyanov and Zamolodchikov to the case of arbitrary, but equal quark masses $m_1=m_2.$ Our results include analytical expressions for spectral sums and systematic large-$n$ expansion. We also study the spectral sums in the chiral limit and the heavy quark limit and find a complete agreement with known results.
\end{abstract}
\section{Introduction}
One of the most difficult and intriguing problems in modern theoretical physics is to understand the mechanism of confinement of colored particles: quarks and gluons. Unfortunately, at the moment it is not possible to analytically solve this problem in strongly coupled Gauge Theories in various dimensions, and in particular in Quantum Chromodynamics (QCD) in dimension $3+1$. One might hope that studying this problem in various toy models, will bring us closer to the solution of this long-standing question.

In this paper, we will consider a $1+1$-dimensional version of multicolor QCD, the so-called QCD$_2$ 't Hooft's model \cite{THOOFT1974461}. This theory is one of the simplest known toy models demonstrating the phenomenon of confinement.  The model is described by the Lagrangian density of $SU(N_c)$ Yang-Mills theory coupled to $N_f$ massive fermions in fundamental representation 
\begin{equation}
    \mathcal{L} =-\frac{1}{2}\text{tr\;}F_{\mu\nu}F^{\mu\nu} +\sum\limits_{k=1}^{N_f}\bar{\psi}_k(i\gamma^\mu D_\mu-m_k)\psi_k.
\end{equation}
We are interested in the double scaling limit at which the number of colors $N_c$ grows, while keeping the 't Hooft's coupling constant $g^2=g^2_{{\text{YM}}}N_c$ fixed. In this limit only planar Feynman diagrams are relevant, and diagrams with self-intersections are suppressed by the power factors $1/N_c$ (depending on the number of self-intersections). As a result, there are no free quarks in this theory, and the spectrum of bound states, the mesons, resembles the Regge trajectories known in QCD$_4$. Therefore, this low-dimensional model turns out to be convenient for studying the mechanisms of confinement.

Since the mesons in this model are composite objects, their masses are precisely determined by the Bethe-Salpeter equation. For mesons that consist of quark and antiquark with the bare (Lagrangian) masses $m_i \;(i=1,2),$ the Bethe-Salpeter equation can be reduced to the integral equation called the 't Hooft equation \cite{THOOFT1974461}
\begin{equation}\label{'tHooft-eq}
    2\pi^2\lambda\;\phi_{12}(x)=\mathcal{H}\phi_{12}(x)=\left[\frac{\alpha_1}{x}+\frac{\alpha_2}{1-x}\right]\phi_{12}(x)-\fint_0^1\limits dy\frac{\phi_{12}(y)}{(x-y)^2},\quad \alpha_i=\frac{\pi m^2_i}{g^2}-1.
\end{equation}
Here $\phi_{12}(x)$ is the wave function of the meson state, which has the following asymptotic behavior at the endpoints of the interval $[0,1]$
\begin{equation}\label{boundary-cond}
    \phi_{12}(x)\sim
    \begin{cases}
        x^{\beta_1},\quad &x\to0;\\
        (1-x)^{\beta_2},\quad &x\to1,
    \end{cases}
\end{equation}
and $\beta_i$ are the zeros of the equation
\begin{equation}
     \pi\beta_i \cot{(\pi\beta_i)}+\alpha_i=0,\quad 0\leq\beta_i<1.
\end{equation}
Equation \eqref{'tHooft-eq} is the spectral problem for the parameter $\lambda.$ The meson masses $M^2_n,\; n=0,1,2, \dots$ are discrete and related to the eigenvalues $\lambda_n$ as
\begin{equation}\label{M-and-lambda}
    M^2_n=2\pi g^2\lambda_n.
\end{equation}

Despite the fact that 't Hooft equation can be solved numerically with high accuracy, and the main characteristics of this theory can be obtained from the numerical solution, the search for analytical solutions is also of interest. In their seminal paper \cite{Fateev:2009jf} Fateev, Lukyanov, and Zamolodchikov (FLZ) initiated the development of analytic methods for 't Hooft equation. Their paper studied the special case of quark masses $\alpha_1 =\alpha_2 = 0$, for which the problem was reformulated in terms of Baxter's TQ equation (see section \ref{From-tHooft-to-TQ}). The authors considered two limiting cases:
\begin{itemize}
    \item low energy levels (small $\lambda,$ perturbation theory);
    \item high energy levels (large $\lambda,$ quasi-classical).
\end{itemize}
They managed to calculate the spectral sums $G^{(s)}_{\pm}$ (for $s\leq 13$)\footnote{Since the leading asymptotic of large eigenvalues is proportional to $n$, according to this definition, the first spectral sums diverge, so they need to be regularized.}
\begin{equation}
\begin{aligned}\label{spectral_sums_def}
    &G^{(1)}_{+}\overset{\text{def}}{=}\sum_{n=0}^{\infty}\left[\frac{1}{\lambda_{2n}}-\frac{1}{n+1}\right]\quad&&\text{and}\quad G^{(1)}_{-}\overset{\text{def}}{=}\sum_{n=0}^{\infty}\left[\frac{1}{\lambda_{2n+1}}-\frac{1}{n+1}\right],\\
    &G^{(s)}_{+}\overset{\text{def}}{=}\sum_{n=0}^{\infty}\frac{1}{\lambda_{2n}^s}\quad&&\text{and}\quad G^{(s)}_{-}\overset{\text{def}}{=}\sum_{n=0}^{\infty}\frac{1}{\lambda_{2n+1}^s}, \quad s=2,3,\dots,
\end{aligned}
\end{equation}
as well as have found a quasi-classical formula for high energy eigenvalues  
\begin{equation}\label{WKB_alpha=0}
    \lambda_n=\frac{1}{2}\left(n+\frac{3}{4}\right)-\frac{1}{3\pi^6(n+\frac{3}{4})^3}+\frac{(-1)^{n+1}}{\pi^4(n+\frac{3}{4})^2}\left\{1-\frac{4\log{(\pi e^{\gamma_E-\frac{1}{2}}}(n+\frac{3}{4}))}{\pi^2(n+\frac{3}{4})}\right\}+\mathcal{O}\left(\frac{\log^2{n}}{n^4}\right).
\end{equation}
They also presented, but did not provide details of the derivation, a generalization of the formula \eqref{WKB_alpha=0} and the analytic expressions for $G^{(1)}_{\pm}$ to the case $\alpha_1=\alpha_2=\alpha$. 

In our work, we take over from the FLZ approach and address the problem of determining the meson spectrum in 't Hooft's model for quarks with equal masses, that is $\alpha_1=\alpha_2=\alpha$. Following \cite{Fateev:2009jf}, we derive Baxter's TQ equation and consider two limiting cases of small and large $\lambda$. We have managed to derive analytic expressions for first few spectral sums, as well as a quasi-classical formula (in a higher order in $1/n$, than provided in \cite{Fateev:2009jf}). In the special case $\alpha\to0$, our results agree with those of FLZ. We also verified our analytical formulas numerically with great accuracy\footnote{We note that the question of generalizing the results of FLZ to the case of $\alpha_1=\alpha_2=\alpha$ has been addressed in the recent paper \cite{Ambrosino:2023dik}. However, we believe that in \cite{Ambrosino:2023dik} the analytic properties of the solutions to the TQ equation have not been correctly identified.}.

The paper is organized as follows. In Section \ref{From-tHooft-to-TQ}, we discuss the integrability view of the equation \eqref{'tHooft-eq} and present our strategy for solving the problem (mainly following FLZ approach). Namely, we derive Baxter's TQ equation and explain how its solutions are related to those of the original integral equation. We also discuss the  relations between these solutions and spectral determinants. In sections \ref{Small-lambda-expansion} and \ref{Large-lambda-expansion} we are concerned with finding solutions to the TQ equation in two different limiting cases of small and large $\lambda,$ respectively. Section \ref{Analytical-results} is devoted to analytical results, among which are expressions for the spectral sums and for the large-$n$ decomposition. In section \ref{Spectral_sums_limits} we study the spectral sums in two physically interesting limiting cases: the chiral limit $\alpha\to-1$ and the heavy quark limit $\alpha\to\infty$. In Section \ref{Numerical-results}, we test our answers by comparing them with the numerical solution of \eqref{'tHooft-eq}, which is based on the decomposition of functions on the basis of Chebyshev polynomials. Finally, in Section \ref{Conclusion} we give concluding remarks and discuss various directions for future work. Several technical details can be found in appendices.
\section{From 't Hooft to Baxter's TQ equation}\label{From-tHooft-to-TQ}
In this section, following \cite{Fateev:2009jf}, we reformulate the problem of finding the meson masses of t Hooft's equation in the language of integrability, namely in terms of Baxter's TQ equation, and explain our strategy for solving it.
\subsection{Fourier form of 't Hooft equation}
It is useful to apply the Fourier transform with the variable $\frac{1}{2}\log{\frac{x}{1-x}}$ to the equation \eqref{'tHooft-eq} (for the first time such a transformation was proposed in \cite{NARAYANAN200576}) 
\begin{equation}
    \phi(x)=\int_{-\infty}^{\infty}\limits\frac{d\nu}{2\pi}\left(\frac{x}{1-x}\right)^{\frac{i\nu}{2}}\Psi(\nu),\quad \Psi(\nu)=\int_0^1\limits\frac{dx}{2x(1-x)}\left(\frac{x}{1-x}\right)^{-\frac{i\nu}{2}}\phi(x).
\end{equation}
The Fourier form of the equation \eqref{'tHooft-eq} is (for $\alpha_1=\alpha_2=\alpha$)
\begin{equation}\label{'tHooft-eq-Fourier}
    \left(\frac{2\alpha}{\pi}+\nu\coth{\frac{\pi\nu}{2}}\right)\Psi(\nu)-\lambda\int^{\infty}_{-\infty}\limits d\nu' \frac{\pi(\nu'-\nu)}{2\sinh{\frac{\pi(\nu'-\nu)}{2}}}\Psi(\nu')=0.
\end{equation}
For the norm $\|\varphi\|^2=\int_{0}^1\limits dx\;|\varphi(x)|^2$ to be finite and the function $\varphi(x)$ in \eqref{'tHooft-eq} to satisfy the boundary conditions \eqref{boundary-cond}, the solution $\Psi(\nu)$ has to decay at $|\nu|\to\infty$, and  be a smooth function of the real variable $\nu$. Careful analysis of the equation \eqref{'tHooft-eq-Fourier} shows that $\Psi(\nu)$ is a meromorphic function of the complex variable $\nu$ with simple poles at 
\begin{equation}
    i\nu_k^*+2iN,\quad -i\nu_k^*-2iN,\quad N\geq0,
\end{equation}
where $\pm i\nu_k^*$ are roots of the equation
\begin{equation}\label{main-trancendental-equation}
    \frac{2\alpha}{\pi}+\nu\coth{\frac{\pi\nu}{2}}=0\quad\text{for}\quad \alpha>-1.
\end{equation}
\subsection{TQ equation}
If we define the new function $Q(\nu)$ by the following expression
\begin{equation}\label{Q_pm-def}
    Q(\nu)\overset{\text{def}}{=}\sinh\frac{\pi\nu}{2}\left(\frac{2\alpha}{\pi}+\nu\coth\frac{\pi\nu}{2}\right)\Psi(\nu),
\end{equation}
we conclude that the analytic properties of the solution $\Psi(\nu)$ imply that $Q(\nu)$ has to satisfy the following requirements
\begin{enumerate}
    \item be analytic in the strip $\textrm{Im}\,\nu\in[-2,2];$
    \item grow slower than any exponential at $|\textrm{Re}\,\nu|\to\infty,$ in other words to be bounded as 
    \begin{equation}\label{Q-is-bounded}
        Q(\nu)=\mathcal{O}(e^{\epsilon|\nu|}),\quad \forall \epsilon>0,\quad |\textrm{Re}\,\nu|\to\infty;
    \end{equation}
    \item obey the quantization conditions
    \begin{equation}\label{quantisation-condition}
        Q(0)=Q(\pm2i)=0.
    \end{equation}
\end{enumerate}
In terms of the $Q$-function, the equation \eqref{'tHooft-eq-Fourier} takes the following form 
\begin{equation}\label{Thooft-Q}
    Q(\nu)=\lambda\sinh{\frac{\pi\nu}{2}}\int_{-\infty}^{\infty}\limits d\nu'\frac{\pi(\nu-\nu')}{2\sinh{\pi(\nu-\nu')}}\frac{Q(\nu')}{\nu'\cosh{\frac{\pi\nu'}{2}}+\frac{2\alpha}{\pi}\sinh{\frac{\pi\nu'}{2}}}.
\end{equation}
In particular, the quantization condition \eqref{quantisation-condition} follows directly from \eqref{Thooft-Q}.

Let us consider the following linear combination of $Q$ functions 
\begin{equation}\label{QQQ-combination}
    Q(\nu+i\epsilon+2i)+Q(\nu+i\epsilon-2i)-2Q(\nu+i\epsilon),
\end{equation}
and use \eqref{Thooft-Q}. We note that for $\epsilon>0$ the r.h.s. of \eqref{Thooft-Q} for $Q(\nu+i\epsilon+2i)$ acquires  additional residue term
\begin{equation}
    \lambda\sinh{\frac{\pi(\nu+2i)}{2}}\varointclockwise_{\mathcal{C}_\nu}d\nu'\frac{\pi(\nu-\nu'+2i)}{2\sinh{\frac{\pi(\nu-\nu'+2i)}{2}}}\frac{Q(\nu')}{\nu'\cosh{\frac{\pi\nu'}{2}}+\frac{2\alpha}{\pi}\sinh{\frac{\pi\nu'}{2}}}=-\frac{4\pi\lambda\sinh{\frac{\pi\nu}{2}}}{\nu\cosh{\frac{\pi\nu}{2}}+\frac{2\alpha}{\pi}\sinh{\frac{\pi\nu}{2}}}Q(\nu),
\end{equation}
while the integral terms cancel exactly for the combination \eqref{QQQ-combination}. Thus we obtain the functional difference equation
\begin{equation}\label{TQ-equation}
    Q(\nu+2i)+Q(\nu-2i)-2Q(\nu)=-\frac{2z}{\nu+\alpha x}Q(\nu),
\end{equation}
where
\begin{equation}\label{xz-expression}
    z=2\pi\lambda \tanh\left(\frac{\pi\nu}{2}\right)\quad\text{and}\quad
    x=\frac{2}{\pi}\tanh\left(\frac{\pi\nu}{2}\right).
\end{equation}
Equation \eqref{TQ-equation} has the form of the famous Baxter's TQ equation and our goal, using the solutions of this equation, is to obtain the analytical expressions the meson masses of the original 't Hooft's equation.
\subsection{Inhomogeneous integral equation}
Before we proceed directly to the search of the solution, let us say few words about the quantization conditions. A solution to \eqref{TQ-equation} with the above three properties exists only at specific values of $\lambda$, which determine the eigenvalues of \eqref{'tHooft-eq-Fourier}. However, if we relax the conditions and drop the last one \eqref{quantisation-condition}, then solutions $Q(\nu|\lambda)$ will exist for any $\lambda$. For arbitrary $\lambda$, the corresponding function $\Psi(\nu|\lambda)$ will no longer satisfy the homogeneous integral equation \eqref{'tHooft-eq-Fourier}, but rather inhomogeneous one\footnote{It can be derived by standard methods of complex analysis. Namely, we apply the operator $\hat{K}$, where
\begin{equation*}
   \hat{K}\cdot f(\nu)=\fint_{-\infty}^{\infty}\limits\frac{\pi(\nu-\nu')}{2\sinh\frac{\pi(\nu-\nu')}{2}}\frac{1}{\sinh\frac{\pi\nu'}{2}}f(\nu')d\nu' 
\end{equation*}
to the both sides of TQ equation \eqref{TQ-equation}. The r.h.s. of \eqref{TQ-equation} will result in the $\lambda$ term of \eqref{'tHooft-eq-Fourier-Inhomogenious}. In the l.h.s. of \eqref{TQ-equation} all three integrals can be reduced to one, which then vanishes identically, picking some residue terms. There are two types of such terms, either coming from the function $\frac{\pi(\nu-\nu')}{2\sinh\frac{\pi(\nu-\nu')}{2}}$ which produce $\left(\frac{2\alpha}{\pi}+\nu\coth{\frac{\pi\nu}{2}}\right)\Psi(\nu)$ term in \eqref{'tHooft-eq-Fourier-Inhomogenious}, or from $\frac{1}{\sinh\frac{\pi\nu'}{2}}$ which produces r.h.s. of \eqref{'tHooft-eq-Fourier-Inhomogenious}. We leave it as an exercise to relate $q_{\pm}(\lambda)$ to the values $Q(0)$ and $Q(\pm2i)$.}
\begin{equation}\label{'tHooft-eq-Fourier-Inhomogenious}
    \left(\frac{2\alpha}{\pi}+\nu\coth{\frac{\pi\nu}{2}}\right)\Psi(\nu)-\lambda\fint^{\infty}_{-\infty}\limits d\nu' \frac{\pi(\nu-\nu')}{2\sinh{\frac{\pi(\nu-\nu')}{2}}}\Psi(\nu')=F(\nu|\lambda),\quad F(\nu|\lambda)=\frac{q_+(\lambda)\nu+q_-(\lambda)}{\sinh{\frac{\pi\nu}{2}}},
\end{equation}
where $q_{\pm}(\lambda)$ are linear combinations of $Q(0)$ and $Q(\pm 2i)$. It can be shown that once the coefficients $q_{\pm}(\lambda)$ are chosen the solution of \eqref{'tHooft-eq-Fourier-Inhomogenious} is unique. Thus, we are dealing with a two-dimensional space of functions $\Psi(\nu|\lambda)$, where we can choose a basis of symmetric and antisymmetric functions
\begin{equation}
    \Psi_{\pm}(-\nu|\lambda)=\pm\Psi_{\pm}(\nu|\lambda),
\end{equation}
corresponding to
\begin{equation}\label{F-choice}
    F_+(\nu|\lambda)=\frac{\nu}{\sinh{\frac{\pi\nu}{2}}}\quad\text{and}\quad F_-(\nu|\lambda)=\frac{1}{\sinh{\frac{\pi\nu}{2}}}
\end{equation}
respectively.

As one already mentioned,  equation \eqref{'tHooft-eq-Fourier-Inhomogenious} has a unique solution for any $\lambda$ for each non-zero fixed r.h.s. But if $\lambda$ approaches the spectral point $\lambda\rightarrow\lambda_n$ then the original homogeneous equation \eqref{'tHooft-eq-Fourier} has to be recovered. It means that the function $ \Psi_{\pm}(\nu|\lambda)$ should diverge at the spectral points. More careful analysis \cite{Fateev:2009jf} shows that $\Psi_{\pm}(\nu|\lambda)$ are meromorphic functions of $\lambda$ with singularities only at $\lambda_n$   
\begin{equation}\label{Psi-analytic-lambda}
    \Psi_+(\nu|\lambda)=\sum_{n=0}^{\infty}\frac{c_{2n}\Psi_{2n}(\nu)}{\lambda-\lambda_{2n}},\quad \Psi_-(\nu|\lambda)=\sum_{n=0}^{\infty}\frac{c_{2n+1}\Psi_{2n+1}(\nu)}{\lambda-\lambda_{2n+1}}.
\end{equation}
The coefficients $c_n$ have a deep physical meaning \cite{Fateev:2009jf,Callan:1976PhysRev}: they are related to the matrix elements of the vector current $J_{\mu}=\bar{\psi}\gamma_\mu\psi$ and the scalar density $\bar{\psi}\psi$.
\subsection{Quantum Wronskian relation}
The symmetric and antisymmetric functions $\Psi_{\pm}(\nu|\lambda)$ are used to construct the functions $Q_{\pm}(\nu|\lambda)$ (further, for simplicity, we will sometimes hide the dependence on $\lambda$, but it will always be implied)
\begin{equation}
    Q_{\pm}(-\nu)=\mp Q_{\pm}(\nu).
\end{equation}
They are two independent solutions of the inhomogeneous integral equation \eqref{'tHooft-eq-Fourier-Inhomogenious} and also of TQ equation \eqref{TQ-equation}. According to the choice of $F_{\pm}(\nu|\lambda)$ in \eqref{F-choice}, these functions have the following normalization conditions
\begin{equation}\label{Q-normalization-cond}
    Q_+(2i)=2i \quad\text{and}\quad Q_-(0)=1.
\end{equation}

Baxter's TQ equation is known to play an important role in quantum integrable systems and has a large set of properties, one of which is associated with the so-called quantum Wronskian
\begin{equation}\label{Quantum-Wronskian}
    W(\nu|\lambda)\overset{\text{def}}{=}Q_+(\nu+i)Q_-(\nu-i)-Q_+(\nu-i)Q_-(\nu+i).
\end{equation}
By construction, this quantity is a $2i$ periodic function (it follows directly from \eqref{TQ-equation})
\begin{equation}
    W(\nu|\lambda)=W(\nu+2i|\lambda).
\end{equation}
Since the functions $Q_{\pm}(\nu)$ are analytic in the strip $\textrm{Im}\,\nu\in[-2,2]$ and grow slower than any exponent \eqref{Q-is-bounded}, then in fact the Wronskian turns out to be an entire and bounded function, and thus is a constant. It is enough to determine the value of the Wronskian at one point, for example $\nu=i$
\begin{equation}
     W(\nu|\lambda)=W(i|\lambda)=Q_+(2i)Q_-(0)=2i,
\end{equation}
which implies in particular (if we substitute $\nu=0$ in \eqref{Quantum-Wronskian})
\begin{equation}\label{QQ=i}
    Q_+(i)Q_-(i)=i.
\end{equation}
Using the analytical dependence of the functions $\Psi_{\pm}(\nu|\lambda)$ on $\lambda$ \eqref{Psi-analytic-lambda}, we obtain that the same behavior holds for the functions $Q_{\pm}(\nu|\lambda)$
\begin{equation}\label{Q-analytic-lambda}
    Q_+(\nu|\lambda)=\sum_{n=0}^{\infty}\frac{c_{2n}Q_{2n}(\nu)}{\lambda-\lambda_{2n}}\sim\frac{1}{\prod_{n=0}^{\infty}\limits(\lambda-\lambda_{2n})},\quad 
    Q_-(\nu|\lambda)=\sum_{n=0}^{\infty}\frac{c_{2n+1}Q_{2n+1}(\nu)}{\lambda-\lambda_{2n+1}}\sim\frac{1}{\prod_{n=0}^{\infty}\limits(\lambda-\lambda_{2n+1})}.
\end{equation}
Using \eqref{Q-analytic-lambda} together with the ratio \eqref{QQ=i}, we obtain that at the point $\nu=i,$ the poles of the function $Q_{-}$ are the zeros of the function $Q_{+}$ and vice versa
\begin{equation}\label{Q(i)-analytic-lambda}
    Q_+(i)\sim\prod_{n=0}^{\infty}\frac{\lambda-\lambda_{2n+1}}{\lambda-\lambda_{2n}}\quad\text{and}\quad Q_-(i)\sim
    \prod_{n=0}^{\infty}
    \frac{\lambda-\lambda_{2n}}{\lambda-\lambda_{2n+1}}.
\end{equation}
\subsection{Spectral determinants}
Let us define the spectral determinants\footnote{We note that in \cite{Fateev:2009jf} the authors also used another definitions of the spectral determinants. Namely (compare (2.18) and (8.19) in \cite{Fateev:2009jf})
\begin{equation}\label{spectral-determinants-definition-2}
    D_+(\lambda)=\prod_{n=0}^\infty\left(1-\frac{\lambda}{\lambda_{2n}}\right)e^{\frac{\lambda}{\lambda_{2n}}},\quad D_-(\lambda)=e^{2\lambda}\prod_{n=0}^\infty\left(1-\frac{\lambda}{\lambda_{2n+1}}\right)e^{\frac{\lambda}{\lambda_{2n+1}}},
\end{equation}
which is equivalent to
\begin{equation}\label{spectral-determinants-definition-2-2}
D_{\pm}(\lambda)=e^{(1\mp1)\lambda}
    \;\exp\left[-\sum_{s=2}^{\infty}s^{-1}G_{\pm}^{(s)}\lambda^s\right].
\end{equation}
For $\alpha=0$ one has \cite{Fateev:2009jf} $G_+^{(1)}=\log(8\pi)-1$, $G_-^{(1)}=\log(8\pi)-3$ and two definitions \eqref{spectral-determinants-definition-1}-\eqref{spectral-determinants-definition-1-2} and \eqref{spectral-determinants-definition-2}-\eqref{spectral-determinants-definition-2-2} coincide.}
\begin{equation}\label{spectral-determinants-definition-1}
    D_+(\lambda)\overset{\text{def}}{=}\left(\frac{8\pi}{e}\right)^{\lambda}\prod_{n=0}^\infty\left(1-\frac{\lambda}{\lambda_{2n}}\right)e^{\frac{\lambda}{n+1}},\quad D_-(\lambda)\overset{\text{def}}{=}\left(\frac{8\pi}{e}\right)^{\lambda}\prod_{n=0}^\infty\left(1-\frac{\lambda}{\lambda_{2n+1}}\right)e^{\frac{\lambda}{n+1}},
\end{equation}
which can be written in terms of the spectral sums \eqref{spectral_sums_def} as 
\begin{equation}\label{spectral-determinants-definition-1-2}
    D_{\pm}(\lambda)=\left(\frac{8\pi}{e}\right)^{\lambda}
    \exp\left[-\sum_{s=1}^{\infty}s^{-1}G_{\pm}^{(s)}\lambda^s\right].
\end{equation}
Using the resolvent and the Liouville-Neumann series formalism, the following relations between the spectral determinants and the $Q$-functions can be derived directly from the integral equation \eqref{'tHooft-eq-Fourier-Inhomogenious}
\begin{align}\label{DD-integral-prod}
    \partial_{\lambda}\log(D_+D_-)&=2-\int_{-\infty}^{\infty}\limits
    \left[\frac{1}{\nu\coth\frac{\pi\nu}{2}+\frac{2\alpha}{\pi}}
    \left(Q_-(\nu)\partial_{\nu}Q_+(\nu)-Q_+(\nu)\partial_{\nu}Q_-(\nu)\right)-\frac{1}{\nu\coth\frac{\pi\nu}{2}}\right]d\nu,
    \\\label{DD-integral-dev}
    \partial_{\lambda}\log\left(\frac{D_+}{D_-}\right)&=-\int_{-\infty}^{\infty}\limits
    \frac{\pi}{\nu\coth\frac{\pi\nu}{2}+\frac{2\alpha}{\pi}}\frac{Q_+(\nu)Q_-(\nu)}{\sinh\pi\nu}d\nu.
\end{align}
We emphasize that \eqref{DD-integral-prod} and \eqref{DD-integral-dev} are closely related to the possibility of representing the resolvent, via inhomogeneous solutions of $\Psi_{\pm}(\nu|\lambda)$ and the integral kernel (see appendix B in \cite{Fateev:2009jf} for details). In \cite{Its:1990MPhysB}, a class of so-called "completely integrable" kernels with properties similar to our kernel are described. Therefore, we expect that relations similar to \eqref{DD-integral-prod}, \eqref{DD-integral-dev} can also be obtained.

Once the solutions $Q_{\pm}(\nu)$ are constructed, one can use \eqref{DD-integral-prod} and \eqref{DD-integral-dev} to extract the spectral data. However, in practice the direct evaluation of the integrals from \eqref{DD-integral-prod}-\eqref{DD-integral-dev} is not very efficient. In particular, in \cite{Fateev:2009jf} the authors have been able to express the spectral sums $G^{(s)}_{\pm}$ for $\alpha=0$ and $s\leq 13$ through the $\zeta$ values. Actual evaluations of the integrals that leads to their results become very cumbersome as $s$ grows. Moreover \eqref{DD-integral-prod}-\eqref{DD-integral-dev} are suited to study the small $\lambda$ behavior. At large $\lambda$ the integral relation \eqref{DD-integral-prod} behaves badly (but not \eqref{DD-integral-dev}, see appendix \ref{check-2.17}) and we were unable to use it to derive the large $n$ expansion of the spectrum.

In \cite{Fateev:2009jf} the authors have guessed a remarkably simple formula 
\begin{equation}\label{2.17}
    (2.17):\quad\partial_{\lambda}\log D_{\pm}(\lambda)=2i\partial_{\nu}\log Q_{\mp}(\nu)\Bigl|_{\nu=i}
    \quad\text{for}\quad \alpha=0,
\end{equation}
which significantly facilitates computations and works well for both small and large $\lambda$ expansions. It has not been proven rigorously, but passed a number of non-trivial tests. In our work we have found the following generalizations of \eqref{2.17} for $\alpha\neq0$ 
\begin{align}
    \label{2.17-new}
    \partial_{\lambda}\log D_{-}(\lambda)+\alpha\left(\frac{7\zeta(3)}{\pi^2}-\frac{1}{2}\left(\mathtt{u}_1(\alpha)+\alpha \mathtt{u}_3(\alpha)\right)\right)&=2i\,\partial_{\nu}\log Q_{+}(\nu)\Bigl|_{\nu=i},
    \\\label{2.17-new-2}
    \partial_{\lambda}\log D_{+}(\lambda)+\alpha\left(\frac{7\zeta(3)}{\pi^2}-\frac{1}{2}\left(\mathtt{u}_1(\alpha)+\alpha \mathtt{u}_3(\alpha)\right)\right)&=2i\left(1-\frac{2\alpha}{\pi^2}\lambda^{-1}\right)\partial_{\nu}\log Q_{-}(\nu)\Bigl|_{\nu=i},
\end{align}
where $\mathtt{u}_{2k-1}(\alpha)$ will be defined in \eqref{u-def}. Again the status of \eqref{2.17-new}-\eqref{2.17-new-2} remains the same: they lack rigorous proof, but passed all tests that we have done. 

Another non-integral type relations have been conjectured in \cite{Fateev:2009jf} 
\begin{equation}\label{Q/Q=D/D}
    \frac{Q_+(i)}{Q_+(2i)}=\frac{1}{2}\frac{D_-(\lambda)}{D_+(\lambda)}\quad\text{and}\quad
    \frac{Q_-(i)}{Q_-(0)}=\frac{D_+(\lambda)}{D_-(\lambda)},
\end{equation}
for $\alpha=0$. These relations must follow from \eqref{Q(i)-analytic-lambda} under certain analytic conditions satisfied by the spectral determinants. We conjecture that \eqref{Q/Q=D/D} hold for $\alpha\neq0$ as well. One can use \eqref{Q/Q=D/D} on an equal basis with \eqref{2.17-new}, \eqref{2.17-new-2}. The advantage of \eqref{Q/Q=D/D} is that it allows to find the ratio of normalizations of $D_+(\lambda)$ and $D_-(\lambda)$. We note that all the relations \eqref{2.17-new}, \eqref{2.17-new-2} and \eqref{Q/Q=D/D} are insensitive  to the choice of normalizations $Q_{\pm}(\nu)$, while the integral relations \eqref{DD-integral-prod} and \eqref{DD-integral-dev} are valid only for the choice \eqref{Q-normalization-cond}. 
\subsection{Strategy}
We follow the strategy suggested in \cite{Fateev:2009jf} for obtaining analytic expressions for meson mass spectrum. We consider two limiting cases of small and large values of the parameter $\lambda$. Within these limits, we look for symmetric and antisymmetric solutions  $Q_{\pm}(\nu)$ to TQ-equation \eqref{TQ-equation} that are analytic and bounded in the strip $\textrm{Im}\,\nu\in[-2,2]$.  In both cases the solutions satisfying all properties  appear as series over the integer powers of $\lambda^s$ ($s$ is positive for the case of small $\lambda$ and negative for large ones). Then we use \eqref{2.17-new}-\eqref{2.17-new-2}, to obtain the spectral sums and large-$n$ expansion for the spectrum. 
\section{Small $\lambda$ expansion}\label{Small-lambda-expansion}
Key observation of \cite{Fateev:2009jf} was that TQ equation \eqref{TQ-equation} is solved by the  following two series related to confluent hypergeometric function (\eqref{Xi-Sigma} appeared in \cite{Fateev:2009jf} for $\alpha=0$)
\begin{equation}\label{Xi-Sigma}
\begin{aligned}
    &\Xi(\nu|\lambda)=\left(\nu+\alpha x\right)\sum_{k=0}^{\infty}\frac{\left(1+\frac{i(\nu+\alpha x)}{2}\right)_k}{k!(k+1)!}(-iz)^k,\\
    &\Sigma(\nu|\lambda)=1+\sum_{k=1}^{\infty}\frac{\left(\frac{i(\nu+\alpha x)}{2}\right)_k}{k!(k-1)!}\left(\psi\Big(k+\frac{i(\nu+\alpha x)}{2}\Big)-\psi(k)-\psi(k+1)+\psi\Big(\frac{1}{2}\Big)\right)(-iz)^k,
\end{aligned}
\end{equation}
where $\psi(x)$ is digamma function and the parameters $x$ and $z$ are given by \eqref{xz-expression}. It is convenient to have in mind that these two series satisfy the relation
\begin{equation}\label{Sigma-Xi-relation}
    e^{\frac{iz}{2}}\Sigma(\nu|\lambda)-e^{-\frac{iz}{2}}\Sigma(-\nu|\lambda)=
    i\pi^2\lambda
    \frac{\tanh\frac{\pi\nu}{2}}
    {\tanh\frac{\pi(\nu+\alpha x)}{2}}
    e^{\frac{iz}{2}}\,\Xi(\nu|\lambda).
\end{equation}

For $\alpha=0$ the authors of \cite{Fateev:2009jf} defined two functions
\begin{equation}\label{Mpm}
    \begin{aligned}
        &M_+(\nu|\lambda)=e^{\frac{iz}{2}}\Xi(\nu|\lambda),\\
        &M_-(\nu|\lambda)=\frac{1}{2}\left(e^{\frac{iz}{2}}\Sigma(\nu|\lambda)+e^{-\frac{iz}{2}}\Sigma(-\nu|\lambda)\right),
    \end{aligned}
\end{equation}
so that
\begin{equation}
    M_{\pm}(-\nu|\lambda)=\mp M_{\pm}(\nu|\lambda).
\end{equation}
None of \eqref{Mpm} provides a solution to \eqref{TQ-equation} with the desired analiticity in the strip $\textrm{Im}\,\nu\in[-2,2]$. The problem comes from the poles $\nu=\pm i$ of the $\tanh\left(\frac{\pi\nu}{2}\right)$ function. Thus the authors suggested to look for solutions in the form
\begin{equation}\label{FLZ-Q-ansatz}
    Q_{\pm}(\nu|\lambda)=A_{\pm}(\tau|\lambda)M_{\pm}(\nu|\lambda)+
    B_{\pm}(\tau|\lambda)zM_{\mp}(\nu|\lambda),
\end{equation}
where $A_{\pm}(\tau|\lambda)$ and $B_{\pm}(\tau|\lambda)$ admit the expansion in the parameter $\lambda$ with the coefficients which are polynomials in 
\begin{equation}
    \tau=\frac{\pi^2}{4}\tanh^2\left(\frac{\pi\nu}{2}\right).
\end{equation}
We note that since $\tau(\nu)$ is $2i$ periodic function, the Q function \eqref{FLZ-Q-ansatz} solves \eqref{TQ-equation} for any choice of $A_{\pm}(\tau|\lambda)$ and $B_{\pm}(\tau|\lambda)$. Moreover, since $A_{\pm}(\tau|\lambda)$ and $B_{\pm}(\tau|\lambda)$ are even functions on $\nu$, they preserve the symmetry properties of  $Q_{\pm}(\nu|\lambda)$. Regarding them as power series at $\lambda\rightarrow0$ with polynomial coefficients, one can adjust them order by order to ensure the required analiticity. 

In fact, the naive ansatz \eqref{FLZ-Q-ansatz} fails to work for $\alpha\neq0$. The reason is that the functions
\begin{equation}
    \psi\Big(k+\frac{i(\nu+\alpha x)}{2}\Big)=
    \psi\Big(k+\frac{i(\nu+\frac{2\alpha}{\pi}\tanh\frac{\pi\nu}{2})}{2}\Big),
\end{equation}
exhibit unwanted branching point at $\nu=\pm i$. In order to resolve this paradox, we notice that $\Sigma(\nu|\lambda)$ solves TQ equation \eqref{TQ-equation} if we replace 
\begin{equation}\label{psi-replacement}
    \psi\Big(k+\frac{i(\nu+\alpha x)}{2}\Big)\rightarrow 
    \uppsi_{\alpha}\big(\nu-2i(k-1)\big), \quad k=1,2,\dots
\end{equation}
where $\uppsi_{\alpha}\big(\nu\big)$ is a function analytic in the strip $\textrm{Im}\,\nu\in[0,2)$ which obeys the functional relation
\begin{equation}\label{Psi-equation}
    \uppsi_{\alpha}\big(\nu+2i\big)=\uppsi_{\alpha}\big(\nu\big)+\frac{2i}{\nu+\alpha x}.
\end{equation}
Such a function is unique up to a constant shift (see appendix \ref{Shift-equation}). For real $\nu$ it is given by the contour integral 
\begin{equation}\label{psi-alpha-integral-representation}
    \uppsi_{\alpha}\big(\nu+i\big)=-\gamma_E-\log 4+\frac{1}{2}\int_{-\infty-i\epsilon}^{\infty-i\epsilon}\limits\frac{1}{t+\frac{2\alpha}{\pi}\tanh\frac{\pi t}{2}}
    \left(\tanh\frac{\pi t}{2}-\tanh\frac{\pi(t-\nu)}{2}\right)dt.
\end{equation}
The constant term in \eqref{psi-alpha-integral-representation} has been chosen such that
\begin{equation}
    \uppsi_{\alpha=0}\big(\nu\big)=\psi\Big(1+\frac{i\nu}{2}\Big).
\end{equation}
We note that the defining relation \eqref{Psi-equation} implies that the function $\uppsi_{\alpha}\big(\nu\big)$ is analytic in the strip $\textrm{Im}\,\nu\in[-2,2]$ everywhere except two points
\begin{equation}
    \nu=2i\quad\text{and}\quad \nu=-i\nu_1^*,
\end{equation}
where $-i\nu_1^*$ is the solutions to \eqref{main-trancendental-equation} closest to the origin in the bottom half plane. However, these poles do not affect our construction, as they are exactly canceled out by the factors $(-iz)^k$ and $\left(1+\frac{i(\nu+\alpha x)}{2}\right)_k$ in \eqref{Xi-Sigma-new} (see below). We also have the relation
\begin{equation}\label{second-psi-relation}
    \uppsi_{\alpha}\big(\nu\big)-\uppsi_{\alpha}\big(-\nu\big)=
    2i\left(\frac{\pi}{2(1+\alpha)}\coth\frac{\pi\nu}{2}-\frac{1}{\nu+\frac{2\alpha}{\pi}\tanh\frac{\pi\nu}{2}}\right),
\end{equation}
which is useful for actual calculations.

Since we are dealing with the poles coming from $\tanh{\frac{\pi\nu}{2}}$ function, we need the expansion of the function $\uppsi_{\alpha}\big(\nu\big)$ at $\nu=\pm i$
\begin{equation}
    \uppsi_{\alpha}\big(\nu\big)=\uppsi_{0}^{\pm}+\uppsi_{1}^{\pm}\cdot(\nu\mp i)+
    \uppsi_{2}^{\pm}\cdot(\nu\mp i)^2+\dots
\end{equation}
We have
\begin{equation}
    \uppsi_{1}^{+}=\frac{i\pi^2}{4(1+\alpha)},\quad
    \uppsi_{3}^{+}=-\frac{i\pi^4}{48(1+\alpha)},\quad
    \uppsi_{5}^{+}=\frac{i\pi^6}{480(1+\alpha)},\quad \uppsi_{7}^{+}=-\frac{17i\pi^8}{80640(1+\alpha)},\quad\dots
\end{equation}
and
\begin{equation}
\begin{gathered}
    \uppsi_{0}^{+}=-\gamma_E-\log 4,\quad
    \uppsi_{2}^{+}=\frac{7\zeta(3)}{4}-\frac{\pi^2\alpha}{8}\mathtt{u}_3(\alpha),\quad
    \uppsi_{4}^{+}=-\frac{31\zeta(5)}{16}+\frac{\pi^4\alpha}{96}(3\mathtt{u}_5(\alpha)-\mathtt{u}_3(\alpha)),\\
    \uppsi_{6}^{+}=\frac{127\zeta(7)}{64}-\frac{\pi^6\alpha}{5760}(45 \mathtt{u}_7(\alpha)-30\mathtt{u}_5(\alpha)+2\mathtt{u}_3(\alpha)),\\
    \uppsi_{8}^{+}=-\frac{511\zeta(9)}{256}+\frac{\pi^8\alpha}{161280}(
    315\mathtt{u}_{9}(\alpha)-315\mathtt{u}_{7}(\alpha)+63\mathtt{u}_{5}(\alpha)-\mathtt{u}_{3}(\alpha)),
\end{gathered}
\end{equation}
etc where
\begin{equation}\label{u-def}
    \mathtt{u}_{2k-1}(\alpha)\overset{\text{def}}{=}\int_{-\infty}^{\infty}\limits\frac{\sinh^2t}{t\cosh^{2k-1} t \big(\alpha\sinh t+t\cosh t\big)}dt.
\end{equation}
For expansion at the  point $-i$ we use \eqref{second-psi-relation}.

With these preparations we redefine our basic solutions \eqref{Xi-Sigma}
\begin{equation}\label{Xi-Sigma-new}
    \begin{aligned}
        &\Xi(\nu|\lambda)\overset{\text{def}}{=}
        \left(\nu+\alpha x\right)\sum_{k=0}^{\infty}
        \frac{\left(1+\frac{i(\nu+\alpha x)}{2}\right)_k}{k!(k+1)!}(-iz)^k,\\
        &\Sigma(\nu|\lambda)\overset{\text{def}}{=}1+\sum_{k=1}^{\infty}
        \frac{\left(\frac{i(\nu+\alpha x)}{2}\right)_k}{k!(k-1)!}
        \left(\uppsi_{\alpha}\big(\nu-2i(k-1)\big)-\psi(k)-\psi(k+1)+
    \psi\Big(\frac{1}{2}\Big)+\log 16\right)(-iz)^k.
    \end{aligned}
\end{equation}
The difference compared to \eqref{Xi-Sigma} is in the function $\Sigma(\nu|\lambda)$. First, we have made the replacement \eqref{psi-replacement}. And second we shifted $\psi\big(\frac{1}{2}\big)\rightarrow \psi\big(\frac{1}{2}\big)+\log 16$, which amounts to $\Sigma(\nu|\lambda)\rightarrow \Sigma(\nu|\lambda)-iz\log 16\,\Xi(\nu|\alpha)$. This shift eliminates all logs in the coefficients of $A_{\pm}(\tau|\lambda)$, $B_{\pm}(\tau|\lambda)$ and the solution acquires more nice form which coincides with FLZ solution at $\alpha=0$\footnote{Thus we assume that this shift was also done in \cite{Fateev:2009jf}. However it was not stated explicitly.}. Using \eqref{second-psi-relation} we derive the relation which replaces \eqref{Sigma-Xi-relation}
\begin{equation}\label{Sigma-Xi-relation-2}
    e^{\frac{iz}{2}}\Sigma(\nu|\lambda)-e^{-\frac{iz}{2}}\Sigma(-\nu|\lambda)=
    \frac{i\pi^2\lambda}{1+\alpha}
    e^{\frac{iz}{2}}\,\Xi(\nu|\lambda).
\end{equation}
We see that both \eqref{Sigma-Xi-relation} and \eqref{Sigma-Xi-relation-2} coincide for $\alpha=0$ as expected, but for $\alpha\neq0$  the relation \eqref{Sigma-Xi-relation} shows explicitly non-analiticity at $\nu=\pm i$.

Having defined the basic solutions \eqref{Xi-Sigma-new}, we look for solutions of TQ equation \eqref{TQ-equation} in the form
\begin{equation}\label{Q-small-lambda-new}
    Q_{\pm}(\nu|\lambda)=A_{\pm}(\tau|\lambda)M_{\pm}(\nu|\lambda)+
    B_{\pm}(\tau|\lambda)zM_{\mp}(\nu|\lambda),
\end{equation}
where $M_{\pm}(\nu|\lambda)$ are given by \eqref{Mpm} and $A_{\pm}(\tau|\lambda)$ and $B_{\pm}(\tau|\lambda)$ admit the expansion
\begin{equation}\label{AB-expansion}
    A_{\pm}(\tau|\lambda)=1+\sum_{s=1}^{\infty}a_{\pm}^{(s)}(\tau)\lambda^s,\quad
    B_{\pm}(\tau|\lambda)=-(1\pm1)\frac{\alpha}{2\pi^2}\lambda^{-1}+\sum_{s=0}^{\infty}b_{\pm}^{(s)}(\tau)\lambda^s,
\end{equation}
with normalization condition $A_{\pm}(0|\lambda)=1$. The functions $a_{\pm}^{(s)}(\tau)$ and $b_{\pm}^{(s)}(\tau)$ are polynomials in $\tau$ of degree $s$ and $s+1$ respectively. They are uniquely determined by the requirement of absence of poles at $\nu=\pm i$. For example
\begin{equation}\label{ab+}
\begin{gathered}
    a_{+}^{(1)}(\tau)=-\frac{8\alpha}{\pi^2}\,\tau,\quad
    a_{+}^{(2)}(\tau)=\left[1-\frac{24\alpha\left(\pi^2-7\alpha\zeta(3)\right)}{\pi^4}
    -\frac{12\alpha^3\mathtt{u}_3(\alpha)}{\pi^2}\right]\tau-\frac{80\alpha^2}{\pi^4}\tau^2,\\
    b_{+}^{(0)}(\tau)=\left[\frac{1}{4}-\frac{2\alpha\left(\pi^2-7\alpha\zeta(3)\right)}{\pi^4}
    -\frac{\alpha^3\mathtt{u}_3(\alpha)}{\pi^2}\right]-\frac{4\alpha^2}{\pi^4}\tau
\end{gathered}
\end{equation}
and
\begin{equation}\label{ab-}
\begin{gathered}
    a_{-}^{(1)}(\tau)=\frac{8\alpha}{\pi^2}\,\tau,\quad
    a_{-}^{(2)}(\tau)=-\left[1+\frac{8\alpha\left(\pi^2-7\alpha\zeta(3)\right)}{\pi^4}
    +\frac{4\alpha^3\mathtt{u}_3(\alpha)}{\pi^2}\right]\tau+\frac{16\alpha^2}{\pi^4}\tau^2,\\
    b_{-}^{(0)}(\tau)=0,\quad b_{-}^{(1)}(\tau)=\left[2-\frac{14\alpha\zeta(3)}{\pi^3}+\alpha^2\mathtt{u}_3(\alpha)\right]+\frac{8\alpha}{\pi^2}\tau.
\end{gathered}
\end{equation}
We have calculated explicitly the polynomials $a_{\pm}^{(s)}(\tau)$ up to $s=7$ and the polynomials $b_{\pm}^{(s)}(\tau)$ up to $s=6$. At $\alpha\to 0$ the coefficients  $a_{\pm}^{(s)}(\tau)$, $b_{\pm}^{(s)}(\tau)$ reduce to those presented in \cite{Fateev:2009jf} if one multiplies our coefficients $b_{\pm}^{(s)}(\tau)$ by a factor of $2$. Proceeding in this way, one can obtain any desired order in the expansion \eqref{AB-expansion}.
\section{Large $\lambda$ expansion}\label{Large-lambda-expansion}
In order to construct the large $\lambda$ expansion of the functions $Q_{\pm}(\nu|\lambda)$ we note,  following \cite{Fateev:2009jf}, that the series 
\begin{equation}\label{S-ansatz}
    S(\nu)=(-\lambda)^{-\frac{i\nu}{2}}S_0(\nu)\sum_{k=0}^{\infty}
    \frac{\left(1+\frac{i(\nu+\alpha x)}{2}\right)_k\left(\frac{i(\nu+\alpha x)}{2}\right)_k}{k!}\,(iz)^{-k}
\end{equation}
solves TQ equation \eqref{TQ-equation} provided that $S_0(\nu)$ satisfies the functional relation
\begin{equation}\label{S0-equation}
    S_0(\nu+2i)=\frac{4\pi \tanh\left(\frac{\pi\nu}{2}\right)}{\nu+\frac{2\alpha}{\pi} \tanh\left(\frac{\pi\nu}{2}\right)}S_0(\nu).
\end{equation}
It is impossible to meet the analyticity requirements for the functions $Q_{\pm}(\nu|\lambda)$ within the ansatz \eqref{S-ansatz} due to the poles of growth order coming from $(iz)^{-k}$. Nevertheless, the series \eqref{S-ansatz} will serve as a building block for the asymptotic expansion  $Q_{\pm}(\nu|\lambda)$ at large negative $\lambda$. 
We note that for large positive $\lambda$ the series \eqref{S-ansatz} shows exponential growth at $\nu\rightarrow\pm\infty$.

We are interested in the solution to \eqref{S0-equation} analytic, bounded and without zeroes in the strip $\textrm{Im}\,\nu\in[0,2]$. The solution with these properties is unique up to normalization (see appendix \ref{Shift-equation} for details)\footnote{We note that for $\alpha=0$, using
\begin{equation}
    \log\left(\frac{2\tanh\left(\frac{\pi\nu}{2}\right)}{\nu}\right)=
    \int_{0}^{\infty}\limits\frac{dt}{t}\left(\frac{e^{-\frac{t}{4}}\cos\frac{t\nu}{2}}{\cosh\frac{t}{4}}-e^{-t}\right)
\end{equation}
one can rewrite $S_0^{\scriptscriptstyle{\alpha=0}}(\nu+i)$ in the form 
\begin{equation}\label{S0-solution}
    S_0^{\scriptscriptstyle{\alpha=0}}(\nu+i)=\left(2\pi\right)^{-\frac{i\nu}{2}}\exp\left[-\frac{i}{2}\int_0^{\infty}\limits
    \frac{dt}{t}\left(\frac{e^{-\frac{t}{4}}\sin\frac{t\nu}{2}}{\sinh\frac{t}{2}\cosh\frac{t}{4}}-\nu e^{-t}\right)\right],
\end{equation}
which can be expressed in terms of the Barnes function \cite{Fateev:2009jf,Brower:1979PhysRevD}
\begin{equation}
    S_0^{\scriptscriptstyle{\alpha=0}}(\nu)=\left(2\pi\right)^{-\frac{1}{2}-\frac{i\nu}{2}}
    \frac{G\left(2+\frac{i\nu}{2}\right)G\left(\frac{1}{2}-\frac{i\nu}{2}\right)}
    {G\left(1-\frac{i\nu}{2}\right)G\left(\frac{3}{2}+\frac{i\nu}{2}\right)}.
\end{equation}
}
\begin{equation}\label{S0-integral-representation}
    S_0(\nu+i)=\exp\left[\frac{i}{4}\int_{-\infty}^{\infty}\limits\log\left(\frac{4\pi\tanh\left(\frac{\pi t}{2}\right)}{t+\frac{2\alpha}{\pi}\tanh\left(\frac{\pi t}{2}\right)}\right)\left(\tanh\frac{\pi(t-\nu)}{2}-\tanh\frac{\pi t}{2}\right)dt\right],\quad
    S_0(i)=1.
\end{equation}
The solution \eqref{S0-integral-representation} is automatically analytic in the strip $\textrm{Im}\,\nu\in[0,2]$. Using \eqref{S0-equation} it can be continued to the bottom strip 
$\textrm{Im}\,\nu\in(-2,0]$
\begin{equation}\label{S0-continuation}
    S_0(\nu-2i)=\frac{\cosh\left(\frac{\pi\nu}{2}\right)(\nu-2i)+\frac{2\alpha}{\pi}\sinh\left(\frac{\pi\nu}{2}\right)}
    {4\pi \sinh\left(\frac{\pi\nu}{2}\right)}
    S_0(\nu).
\end{equation}
From \eqref{S0-continuation} we conclude that $S_0(\nu)$ is analytic in the entire strip $\textrm{Im}\,\nu\in[-2,2]$ \emph{except} one point $\nu=-2i$ where it has a simple pole. Similar analysis shows the following  meromorphic structure of $S_0(\nu)$
\begin{equation}\label{S-zeroes-poles}
    \begin{aligned}
        &\text{zeroes of order $k$ at:}\quad&&\nu=2(k+1)i,\quad &&&k=1,2,\dots,\\
        &\text{zeroes of order $1$ at:}\quad&&\nu=-i\nu_j^{*}-2ki,\quad&&&k=0,1,\dots,\\
        &\text{poles of order $1$ at:}\quad&&\nu=i\nu_j^{*}+2ki,\quad &&&k=1,2,\dots,\\
        &\text{poles of order $k$ at:}\quad&&\nu=-2ki,\quad &&&k=1,2,\dots,
    \end{aligned}
\end{equation}
where $\pm i\nu_j^*$ are the solutions of the transcendental equation \eqref{main-trancendental-equation}. We note that using the integral representation \eqref{S0-integral-representation} one can derive the following useful relation (see appendix \ref{check-2.17} where this relation is used)
\begin{equation}\label{S0-generic-relation}
    S_0(\nu)S_0(-\nu)=
    \frac{\nu\coth\left(\frac{\pi\nu}{2}\right)+\frac{2\alpha}{\pi}}
    {4\pi}.
\end{equation}

In the analysis below we will need the expansion of $S_0(\nu)$ at $\nu=0,\pm i,\pm2i$. We note that due to the relation \eqref{S0-equation} it is enough to consider the expansion at $\nu=0$ and $\nu=i$
\begin{equation}
    \log S_0(\nu)=\sum_{k=0}^{\infty}s_k(\alpha)\nu^k,\quad
    \log S_0(\nu)=\sum_{k=0}^{\infty}t_k(\alpha)(\nu-i)^k.
\end{equation}
In the case of the expansion at $\nu=0$ the pole $\nu+i$ hits the integral in \eqref{S0-integral-representation} from above. One can notice that in the expansion
\begin{equation}\label{tan-tan-expansion}
    \tanh\frac{\pi(t-\nu)}{2}-\tanh\frac{\pi t}{2}=
    \frac{2}{\sinh \pi t}+\frac{\pi}{\cosh\pi t-1}(\nu+i)+
    \frac{\pi^2\sinh\pi t}{8\sinh^{4}\frac{\pi t}{2}}(\nu+i)^2+\dots
\end{equation}
the coefficients in front of $(\nu+i)^{2k}$ are odd functions of $t$ and hence only half of the residue will contribute.  Thus the coefficients $s_{2k}(\alpha)$ are elementary functions of $\alpha$. For example
\begin{equation}
    s_0(\alpha)=-\frac{1}{2}\log\left(\frac{2\pi^2}{1+\alpha}\right),\quad
    s_2(\alpha)=\frac{\pi^2}{24(1+\alpha)},\quad
    s_4(\alpha)=-\frac{\pi^4}{2880}\frac{7+2\alpha}{(1+\alpha)^2}.
\end{equation}
Contrary, odd coefficients $s_{2k-1}(\alpha)$ correspond to even functions in \eqref{tan-tan-expansion} and hence do not acquire pole contribution, but rather retain the integral part. In particular, we find
\begin{equation}
\begin{aligned}
  &s_1(\alpha)=-\frac{i}{2}(1+\log 2\pi+\gamma_E) +\frac{i\alpha}{8}\mathtt{i}_1(\alpha),\\
  &s_3(\alpha)=\frac{i}{72}\left(2\pi^2+3\zeta(3)\right)+\frac{i\pi^2\alpha}{96}
  \mathtt{i}_3(\alpha)
  ,\\
  &s_5(\alpha)=-\frac{i}{7200}\left(14\pi^4+45\zeta(5)\right)+\frac{i\pi^4\alpha}{1920}
  \left(3\mathtt{i}_5(\alpha)+\mathtt{i}_3(\alpha)\right)
  ,\\
  &s_7(\alpha)=\frac{i}{846720}\left(124\pi^6+945\zeta (7)\right)+
  \frac{i \pi ^6 \alpha}{161280}\left(45\mathtt{i}_7(\alpha)+30\mathtt{i}_5(\alpha)+2\mathtt{i}_3(\alpha)\right),
\end{aligned}
\end{equation}
where
\begin{equation}\label{i2k1-def}
    \mathtt{i}_{2k-1}(\alpha)\overset{\text{def}}{=}\fint_{-\infty}^{\infty}\limits\frac{\sinh2t-2t}{t\sinh^{2k-1} t \big(\alpha\sinh t+t\cosh t\big)}dt.
\end{equation}

Similar analysis shows that $t_{2k}(\alpha)=0$ and
\begin{equation}
    \begin{aligned}
        &t_1(\alpha)=-\frac{i}{2}\left(\gamma_E-1+\log 8\pi\right)+\frac{i\alpha}{8}
        \mathtt{i}_2(\alpha),\\
        &t_3(\alpha)=-\frac{i}{72}\left(2\pi^2-21\zeta(3)\right)-\frac{i\pi^2\alpha}{96}
        \mathtt{i}_4(\alpha),\\
        &t_5(\alpha)=\frac{i}{7200}\left(14\pi^4-1395\zeta(5)\right)+\frac{i\pi^4\alpha}{1920}
        \left(3\mathtt{i}_6(\alpha)-\mathtt{i}_4(\alpha)\right),\\
        &t_7(\alpha)=-\frac{i}{846720}(124\pi^6-120015\zeta(7))-\frac{i\pi^6\alpha}{161280}
        \left(45\mathtt{i}_8(\alpha)-30\mathtt{i}_6(\alpha)+2\mathtt{i}_4(\alpha)\right),
    \end{aligned}
\end{equation}
etc where\footnote{We note that there is the relation between the integrals $\mathtt{i}_{2k}(\alpha)$ and $\mathtt{u}_{2k-1}(\alpha)$ that appear for small $\lambda$ expansion and are defined in \eqref{u-def}. Namely,
\begin{equation}\label{i-u-relation}
    \mathtt{i}_{2k}(\alpha)=2\mathtt{u}_{2k-1}(\alpha)+2\alpha \mathtt{u}_{2k+1}(\alpha)-2\int_{-\infty}^{\infty}\limits\frac{\sinh t}{t \cosh^{2k+1}t}dt.
\end{equation}}
\begin{equation}\label{i2k-def}
    \mathtt{i}_{2k}(\alpha)\overset{\text{def}}{=}\int_{-\infty}^{\infty}\limits\frac{\sinh t(\sinh2t-2t)}{t\cosh^{2k} t \big(\alpha\sinh t+t\cosh t\big)}dt.
\end{equation}

Following \cite{Fateev:2009jf} we represent the solution of TQ equation \eqref{TQ-equation} in the form
\begin{equation}\label{Q-ansatz}
    Q_{\pm}(\nu|\lambda)=T(c^{-1}|\lambda)R_{\pm}(c|\lambda)S(\nu)\mp
    T(-c^{-1}|\lambda)R_{\pm}(-c|\lambda)S(-\nu),
\end{equation}
where $S(\nu)$ is given by \eqref{S-ansatz} and we use the notation
\begin{equation}
c=i\pi\coth\left(\frac{\pi\nu}{2}\right).
\end{equation}
We note that since $c$ is $2i$ periodic function, the ansatz \eqref{Q-ansatz} solves \eqref{TQ-equation} for any $T(c^{-1}|\lambda)$ and $R_{\pm}(c|\lambda)$. The role of these auxiliary functions is to cancel poles of growing order at $\nu=0$, $\nu=\pm2i$ and $\nu=\pm i$ at each order in $\lambda^{-1}$. We look for the functions $T(c^{-1}|\lambda)$ and $R_{\pm}(c|\lambda)$ in the form of asymptotic expansion at large $\lambda$ 
\begin{equation}
    T(c^{-1}|\lambda)=1+\sum_{k=1}^{\infty}T^{(k)}(c^{-1})\lambda^{-k}\quad
    R_{\pm}(c|\lambda)=1+\sum_{k=1}^{\infty}
    R_{\pm}^{(k)}\big(c|\log(-\lambda)\big)\lambda^{-k},
\end{equation}
where $T^{(k)}(c^{-1})$ and $R_{\pm}^{(k)}\big(c|\log(-\lambda)\big)$ are polynomials in their variables. The presence of $\log{(-\lambda)}$ as arguments of the polynomials $R^{(k)}_{\pm}$ is not surprising. It is due to the fact that when looking for the polynomials we have to decompose the function $S(\nu)$ near its poles, as result logarithms will appear due to the factor $(-\lambda)^{-\frac{i\nu}{2}}$.
Moreover, the function $R_{\pm}(c|\lambda)$ is responsible for cancellation of poles at  $\nu=0$ and $\nu=\pm2i$, while $T(c^{-1}|\lambda)$ for poles at $\nu=\pm i$. We have found the closed form expression for the function $T(y|\lambda)$
\begin{equation}
    T(y|\lambda)=\exp{\left[\alpha f\left(\frac{\alpha}{\pi^2\lambda}\right)y\right]},
\end{equation}
where
\begin{equation*}
    f(t)=\frac{\sqrt{1-2t}-1+t(1+\log 4)-2t\log(1+\sqrt{1-2t})}{t}=\frac{t}{2}+
    \frac{t^2}{4}+\frac{5t^3}{24}+\frac{7t^4}{32}+\dots
\end{equation*} 
Unfortunately, we were unable to find similar explicit form of the  functions $R_{\pm}(c|\lambda)$. However, the cancellation of poles at each order in $\lambda^{-1}$ fixes the coefficients  $R_{\pm}^{(k)}\big(c|\log(-\lambda)\big)$  uniquely once the normalization of $Q_{\pm}(\nu|\lambda)$ is fixed. A convenient choice of normalization corresponds to $R_{\pm}(0|\lambda)=1$. Then
\begin{equation}
\begin{gathered}
    R_{\pm}(c|\lambda)=1\pm\frac{(1+\alpha)c}{4\pi^4}\lambda^{-2}
    \pm\frac{(1+\alpha)c\left(6c+6q_1\mp(1+\alpha)\right)}{24\pi^6}\lambda^{-3}+\dots,\\
    q_1=3(1+\alpha)-2\log(-\lambda)-4is_1(\alpha).
\end{gathered}
\end{equation}
We have computed the coefficients $R_{\pm}^{(k)}\big(c|\log(-\lambda)\big)$ for $k\leq8$.

Having at hand the asymptotic form of the functions $Q_\pm(\nu|\lambda)$ for negative large values of $\lambda$,  and using \eqref{2.17-new} and \eqref{2.17-new-2}, we can obtain the asymptotic expansion of the spectral determinants $D_\pm(\lambda)$
\begin{equation}\label{F_definition}
    D_{\pm}(\lambda)=d_{\pm}\left(8\pi e^{-2+\gamma_E}\right)^{\lambda}(-\lambda)^{\lambda-\frac{1}{8}\pm\frac{1}{4}}
    \exp\Bigl(F^{(0)}_{\pm}(L)+F^{(1)}_{\pm}(L)\lambda^{-1}+F^{(2)}_{\pm}(L)\lambda^{-2}+\dots\Bigr),
\end{equation}
where $F_{\pm}^{(k)}(L)$ are some polynomials in $L=\log(-2\pi\lambda)+\gamma_E$. For example 
\begin{equation}
\begin{aligned}
    &F^{(0)}_{\pm}(L)=-\frac{\alpha L^2}{2\pi^2}-\frac{(8\alpha\log{2}-\alpha^2\mathtt{i}_2(\alpha))L}{4\pi^2},\\ 
    &F_{\pm}^{(1)}(L)=\frac{\alpha^2L}{2\pi^4}+\frac{\alpha\pi^2+4\alpha^2(\log{16}-1)-2\alpha^3\mathtt{i}_2(\alpha)\mp\pi^2(4+8\alpha)}{16\pi^4}.
 \end{aligned}   
\end{equation}
The numerical factors $d_{\pm}$ in \eqref{F_definition} can not be derived from \eqref{2.17-new}-\eqref{2.17-new-2}, but their ratio follows from \eqref{Q/Q=D/D}
\begin{equation}\label{dm/dp}
    \frac{d_-}{d_+}=\frac{\sqrt{2(1+\alpha)}}{\pi}.
\end{equation}
This relation is one of the nontrivial predictions of our theory. In the case of $\alpha\to0$ it reproduces the result of \cite{Fateev:2009jf}. The ratio \eqref{dm/dp} can be written as an (fast convergent) infinite product \cite{Fateev:2009jf} 
\begin{equation}\label{dprod}
    \frac{d_-}{d_+}=\frac{\Gamma({\frac{7}{8}})}{\Gamma({\frac{3}{8}})}\left(\prod_{m=0}^{\infty}\limits\,\frac{m+\frac{7}{8}}{\lambda_{2m+1}}\right)\cdot\left(\prod_{m=0}^{\infty}\limits\,\frac{m+\frac{3}{8}}{\lambda_{2m}}\right)^{-1}.
\end{equation}
In section \ref{Numerical-results} we check numerically that this product is equal to \eqref{dm/dp}. 

We have constructed the expansion of the spectral determinants $D_{\pm}(\lambda)$ for $\lambda\to-\infty$. However, the physical values of the meson masses belong to the sector of positive values of $\lambda$. In order to obtain a functional expression for the spectral determinant valid for the physical region of the $\lambda$ plane,  we need to perform an analytic continuation of  $D_{\pm}(\lambda)$ from negative to positive values of $\lambda$. However, we cannot do this naively by continuing through the upper $D_{\pm}(-e^{-i\pi}\lambda)$ or through the lower $D_{\pm}(-e^{+i\pi}\lambda)$ half-plane. One of the brilliant ideas of \cite{Fateev:2009jf} was to analytically continue $D_{\pm}(\lambda)$ as a linear combination  (this way of continuation guarantees the decay of $Q_{\pm}(\nu)$ at $|\rm{Re}\,\nu|\to\infty$)
\begin{equation}
    \mathcal{D}_{\pm}(\lambda)=\frac{1}{2}
    \left(D_{\pm}(-e^{-i\pi}\lambda)+D_{\pm}(-e^{+i\pi}\lambda)\right).
\end{equation}
We have
\begin{equation}\label{D-analytical-continuation}
    \mathcal{D}_{\pm}(\lambda)=2d_{\pm}
    \left(8\pi e^{-2+\gamma_E}\right)^{\lambda}
    \lambda^{\lambda-\frac{1}{8}\pm\frac{1}{4}}\exp{\left(\sum_{k=0}^{\infty}\limits\Xi^{(k)}_{\pm}(l)\lambda^{-k}\right)}\cos\left(
    \frac{\pi}{2}\left[2\lambda-\frac{1}{4}\pm\frac{1}{2}+\sum_{k=0}^{\infty}\Phi^{(k)}_{\pm}(l)\lambda^{-k}\right]\right),
\end{equation}
where the coefficients $\Xi^{(k)}_{\pm}(l)$ and $\Phi^{(k)}_{\pm}(l)$ are polynomials in
\begin{equation}
l=\log{(2\pi\lambda)}+\gamma_E.    
\end{equation}
They are related to the polynomials $F^{(k)}_{\pm}(L)$ in a simple way
\begin{equation}
    \Xi^{(k)}_{\pm}(l)\overset{\text{def}}{=}\frac{1}{2}\left(F^{(k)}_{\pm}(l+i\pi)+F^{(k)}_{\pm}(l-i\pi)\right),\quad
    \Phi^{(k)}_{\pm}(l)\overset{\text{def}}{=}\frac{i}{\pi}\left(F^{(k)}_{\pm}(l-i\pi)-F^{(k)}_{\pm}(l+i\pi)\right).
\end{equation}

By definition of the spectral determinants, their zeros are the meson masses (for $\mathcal{D}_+(\lambda)$ even eigenvalues of $\lambda_{2m}$ and for $\mathcal{D}_-(\lambda)$ odd $\lambda_{2m+1}$). Since by construction the functions $\Xi^{(k)}_{\pm}(\lambda)$, $\Phi^{(k)}_{\pm}(\lambda)$ are real, the expression \eqref{D-analytical-continuation} vanishes when the quantization conditions are met
\begin{equation}\label{quantisation-condition-lambda}
    2\lambda-\frac{1}{4}\pm\frac{1}{2}+\Phi^{(0)}_{\pm}(l)+\Phi^{(1)}_{\pm}(l)\lambda^{-1}+\dots=2m+1,\quad m=0,1,2,\dots
\end{equation}
Thus $\Phi^{(s)}_{\pm}(l)$ are responsible for quantization condition. The first five values of $\Phi^{(s)}_{\pm}(l)$ are\footnote{In \cite{Fateev:2009jf} the polynomials $\Phi^{(s)}_{\pm}(l)$ were given for $s\leq 3$ (see (6.1) in \cite{Fateev:2009jf}). We have found an agreement with the results of \cite{Fateev:2009jf} except $\Phi^{(3)}_{\pm}(l)$ were the coefficient in front of the integral $\mathtt{i}_1(\alpha)$ was $\frac{\alpha}{8}$. In our case it is $\frac{\alpha}{4}$.}
\begin{equation}
\resizebox{\textwidth}{!}{$
\begin{aligned}
    &\Phi^{(0)}_{\pm}(l)=-\frac{\alpha(4\log{4}-\alpha \mathtt{i}_2(\alpha))}{2\pi^2}-\frac{2\alpha}{\pi^2}l,\quad \Phi^{(1)}_{\pm}(l)=\frac{\alpha^2}{\pi^4},\quad \Phi^{(2)}_{\pm}(l)=\frac{\alpha^3\pm\pi^2(1+\alpha)}{2\pi^6},
    \\&\Phi^{(3)}_{\pm}(l)=\frac{1}{12\pi^8}\left[5\alpha^4+\pi^2(1+\alpha)^2 \mp12\pi^2(1+\alpha)\left(l-\frac{1}{2}-\frac{3\alpha}{2}-\frac{\alpha}{4}\mathtt{i}_1(\alpha)\right)\right],
    \\&\Phi^{(4)}_{\pm}(l)=-\frac{(1+\alpha)^2}{4\pi^8}l+\frac{7\alpha^5+\pi^2(1+\alpha)^2(1+5\alpha)+\pi^2(1+\alpha)^2\alpha\mathtt{i}_1(\alpha)}{16\pi^{10}}\mp\Biggl[-\frac{3(1+\alpha)}{2\pi^8}l^2+
    \\&+\frac{(1+\alpha)(22\alpha+3\alpha\mathtt{i}_1(\alpha)+10)}{4\pi^8}l+\frac{8\pi^2(4+3\alpha)-(1+\alpha)(8+24\alpha(4+5\alpha)-(20+44\alpha+3\alpha\mathtt{i}_1(\alpha))\alpha\mathtt{i}_1(\alpha))}{32\pi^8}\Biggl].
\end{aligned}$}
\end{equation}
More values $\Phi^{(s)}_{\pm}(l)$ for $s\leq7$ can be found in \texttt{Phi.nb}.
The large-$n$ expansion follows directly from \eqref{quantisation-condition-lambda} (see Subsection \ref{Large-n-expansion}).  
\section{Analytical results}\label{Analytical-results}
\subsection{Spectral sums} 
In principle, one can obtain the spectral sums by direct evaluation of the integrals in \eqref{DD-integral-prod} and \eqref{DD-integral-dev}. For example, in order to find $G^{(1)}_{\pm}$ it is enough to know only the leading asymptotic of the $Q_{\pm}(\nu|\lambda)$. Using \eqref{Q-small-lambda-new}-\eqref{AB-expansion} we find
\begin{equation}
    Q_-(\nu|\lambda)=1+\mathcal{O}(\lambda),\quad 
    Q_+(\nu|\lambda)=\nu+\mathcal{O}(\lambda).
\end{equation}
Plugging the leading asymptotic into \eqref{DD-integral-prod}-\eqref{DD-integral-dev}, we obtain
\begin{equation}
    \begin{aligned}
        &\partial_{\lambda}\log\left(D_+D_-\right)\Big|_{\lambda=0}=2\log{(8\pi)}-2-G^{(1)}_{+}-G^{(1)}_{-}=2-\int_{-\infty}^{\infty}\limits
        \left[\frac{1}{\nu\coth\frac{\pi\nu}{2}+\frac{2\alpha}{\pi}}-\frac{1}{\nu\coth\frac{\pi\nu}{2}}\right]d\nu,
        \\
        &\partial_{\lambda}\log\left(\frac{D_+}{D_-}\right)\Big|_{\lambda=0}=G^{(1)}_{-}-G^{(1)}_{+}=-\int_{-\infty}^{\infty}\limits
        \frac{\pi}{\nu\coth\frac{\pi\nu}{2}+\frac{2\alpha}{\pi}}\frac{\nu}{\sinh\pi\nu}d\nu.
    \end{aligned}
\end{equation}
The integrals in the right-hand sides by simple manipulations can be expressed in terms of $\mathtt{u}_{2k-1}(\alpha)$ given by \eqref{u-def}. As a result we obtain
\begin{equation}\label{G1-analytical}
    \begin{aligned}
        &G_{+}^{(1)}=\log(8\pi)-1-\frac{7\alpha\zeta(3)}{\pi^2}-\frac{\alpha}{2}\left(\mathtt{u}_1(\alpha)-\alpha \mathtt{u}_3(\alpha)\right),\\
        &G_{-}^{(1)}=\log(8\pi)-3+\frac{7\alpha\zeta(3)}{\pi^2}-\frac{\alpha}{2}\left(\mathtt{u}_1(\alpha)+\alpha\mathtt{u}_3(\alpha)\right).
    \end{aligned}
\end{equation}

Proceeding the same way, one can compute higher spectral sums as well. However, this is not a simple task as multiple integrals will appear. In order to facilitate the calculation, we will use \eqref{2.17-new} and \eqref{2.17-new-2}.  As a result, we obtain the spectral sums $G^{(s)}_{\pm}$ with $s>1$:
\begin{equation}\label{G2-analytical}
\begin{aligned}
    &G^{(2)}_{+}=7\zeta(3)+8\alpha\left[\frac{1}{3}-\frac{7\zeta(3)}{\pi^2}\right]+\frac{4\alpha^2}{\pi^2}\left[-\frac{28\zeta(3)}{3}+\frac{49\zeta^2(3)}{\pi^2}+\frac{62\zeta(5)}{\pi^2}\right]+
    \\&+\left[-\frac{\pi^2}{2}+4\alpha+4\alpha^2-\frac{28\alpha^2\zeta(3)}{\pi^2}\right]\alpha\mathtt{u}_3(\alpha)+\alpha^4\mathtt{u}_3^2(\alpha)-4\alpha^3\mathtt{u}_5(\alpha),
    \\&G^{(2)}_{-}=2-\frac{4\alpha}{3}+\frac{4\alpha^2}{\pi^2}\left[\frac{14}{3}\zeta(3)-\frac{31}{\pi^2}\zeta(5)\right]-2\alpha^3\mathtt{u}_3(\alpha)+2\alpha^3\mathtt{u}_5(\alpha),
\end{aligned}
\end{equation}
\begin{equation}\label{G3-analytical}
\resizebox{\textwidth}{!}{$
\begin{aligned}
    &G^{(3)}_{+}=-\frac{4\pi^2}{3}+28\zeta(3)+\alpha\left[4+\frac{70\zeta(3)}{3}-\frac{84\zeta(3)+147\zeta^2(3)+155\zeta(5)}{\pi^2}\right]+
    \\&+12\alpha^2\left[\frac{1}{5}-\frac{14\zeta(3)}{\pi^2}+\frac{2\left(49\zeta^2(3)+31\zeta(5)\right)}{\pi^4}\right]-
    \\&-\frac{4\alpha^3}{\pi^2}\left[\frac{161\zeta(3)}{15}-\frac{4(49\zeta^2(3)+31\zeta(5))}{\pi^2}+\frac{14\zeta(3)(49\zeta^2(3)+93\zeta(5))+5715\zeta(7)}{\pi^4}\right]-
    \\&-\left[2\pi^2-\alpha\left(6-\frac{5\pi^2}{2}+21\zeta(3)\right)-8\alpha^2\left(2-\frac{21\zeta(3)}{\pi^2}\right)-2\alpha^3\left(3-\frac{70\zeta(3)}{\pi^2}+\frac{294\zeta^2(3)+186\zeta(5)}{\pi^4}\right)\right]\alpha \mathtt{u}_3(\alpha)+
    \\&+3\left[-\frac{\pi^2}{4}+2\alpha+2\alpha^2\left(1-\frac{7\zeta(3)}{\pi^2}\right)\right]\alpha^3\mathtt{u}_3^2(\alpha)+\alpha^6\mathtt{u}_3^3(\alpha)+\left[\frac{5\pi^2}{2}-12\alpha+12\alpha^2\left(\frac{7\zeta(3)}{\pi^2}-1\right)\right]\alpha^2\mathtt{u}_5(\alpha)-
    \\&-6\alpha^5\mathtt{u}_3(\alpha)\mathtt{u}_5(\alpha)+6\alpha^4\mathtt{u}_7(\alpha),
    \\&G^{(3)}_{-}=\frac{4\pi^2}{9}-\frac{8}{3}+2\alpha\left[\frac{4}{3}-\frac{14\zeta(3)}{3}+\frac{31\zeta(5)}{\pi^2}\right]-\frac{8\alpha^2}{5}+\frac{8\alpha^3}{\pi^2}\left[\frac{161\zeta(3)}{45}-\frac{124\zeta(5)}{3\pi^2}+\frac{127\zeta(7)}{\pi^4}\right]+
    \\&+[\pi^2-4\alpha^2]\alpha^2\mathtt{u}_3(\alpha)-[\pi^2-8\alpha^2]\alpha^2\mathtt{u}_5(\alpha)-4\alpha^4\mathtt{u}_7(\alpha),
\end{aligned}$}
\end{equation}
which in the case $\alpha=0$ coincide with \cite{Fateev:2009jf}. 

The analytic expressions for $G_{\pm}^{(1)}$ were already presented in \cite{Fateev:2009jf}. The results for $G_{\pm}^{(s)}$ with $s>1$ are new. We collect more spectral sums in \texttt{spectral-sums.nb} (up to $s=7$). Note that analytic expressions for $G_{\pm}^{(s)}$ with $s>1$ depend on the validity of \eqref{2.17-new} and \eqref{2.17-new-2} and thus are of conjectural nature. We have verified them numerically with high accuracy using the Chebyshev polynomial method as in \cite{Fateev:2009jf} (see section \ref{Numerical-results}).
\subsection{Large -- $n$ expansion}\label{Large-n-expansion}
Using the quantization condition \eqref{quantisation-condition-lambda}, one can derive the large $n$-expansion of the eigenvalues $\lambda_n$
\begin{multline}\label{WKB-eq}
    \lambda_n=\frac{1}{2}\mathfrak{n}+\frac{\alpha}{\pi^2}\log{\rho}+\frac{\alpha^2}{\pi^4}\frac{2\log\rho-1}{\mathfrak{n}}-\frac{1}{\pi^4}\frac{1}{\mathfrak{n}^2}\Biggl[\frac{2\alpha^3}{\pi^2}\log^2{\rho}-\frac{6\alpha^3}{\pi^2}\log\rho+\frac{3\alpha^3}{\pi^2}+(-1)^n(1+\alpha)\Biggl]+
    \\+\frac{1}{\pi^6}\frac{1}{\mathfrak{n}^3}\Biggl[\frac{8\alpha^4}{3\pi^2}\log^3\rho-\frac{16\alpha^4}{\pi^2}\log^2\rho+\frac{24\alpha^4}{\pi^2}\log\rho-\frac{29\alpha^4+\pi^2(1+\alpha)^2}{3\pi^2}+
    \\+(-1)^n(1+\alpha)\left(4(1+\alpha)\log\rho-(2+8\alpha+8\log{2}+\alpha\mathtt{i}_1(\alpha))\right)\Biggl]+\mathcal{O}\left(\frac{\log^4{\mathfrak{n}}}{\mathfrak{n}^4}\right),
\end{multline}
where
\begin{equation}
    \mathfrak{n}=n+\frac{3}{4}-\frac{\alpha^2}{2\pi^2}\mathtt{i}_2(\alpha),\quad\rho=4\pi e^{\gamma_E}\left(n+\frac{3}{4}-\frac{\alpha^2}{2\pi^2}\mathtt{i}_2(\alpha)\right).
\end{equation}
In fact the result \eqref{WKB-eq} up to the order $\frac{1}{\mathfrak{n}^3}$ follows from the "quantization condition" announced in \cite{Fateev:2009jf} (formula (6.1)). Higher-order terms in \eqref{WKB-eq} can be computed from \eqref{quantisation-condition-lambda} in a regular manner provided that the coefficients $\Phi^{(s)}_{\pm}(l)$ are known. However, this exercise becomes tedious as the order grows. We explicitly did it up to $\frac{1}{\mathfrak{n}^6}$ (see Mathematica notebook \texttt{WKB.nb} attached to this submission).

We note that the use of \eqref{2.17-new} and \eqref{2.17-new-2} in deriving \eqref{quantisation-condition-lambda} and \eqref{WKB-eq} can not be avoided. The use of integral identities \eqref{DD-integral-prod} and \eqref{DD-integral-dev} is potentially problematic since they were derived as small $\lambda$ expansion and analytic continuation to the large $\lambda$ region might not exist. Although the integral identity \eqref{DD-integral-dev} works well for large $\lambda$ expansion and can be used in particular as an analytic check of \eqref{2.17-new}-\eqref{2.17-new-2} (see appendix \ref{check-2.17}), the integral in \eqref{DD-integral-prod} seems to diverge, and we were unable to find any reasonable way of using it.
\section{Spectral sums in limiting cases}\label{Spectral_sums_limits}
In this section we obtain asymptotics for the spectral sums obtained in section \eqref{Analytical-results} in two limiting cases:
\begin{itemize}
    \item the chiral limit $\alpha\to-1$;
    \item the heavy quark mass limit $\alpha\to\infty$.
\end{itemize}
These special cases allow us to compare our expressions with the results of other works. In the following, we will use the asymptotics of the integrals $\mathtt{u}_{2k-1}(\alpha)$ derived in Appendix \ref{u-asymptotics}.
\subsection{The chiral limit and the ground state}
From a physical point of view, the chiral limit is of great interest. In \cite{Zhitnitsky:1985um}, using QCD-sum rule technique, an exact result for the condensate in the chiral limit $m\to0$ in the 't Hooft model was derived, and later in \cite{Burkardt:1995eb} a formula for arbitrary quark mass was obtained
\begin{equation}
    \braket{\bar\psi\psi}=\frac{mN_c}{2\pi}\left\{\log{\frac{\pi}{1+\alpha}}-1-\gamma_E+\alpha\log4-\alpha^2 I(\alpha)\right\},\quad I(\alpha)=\frac{1}{4}\int^{\infty}_{-\infty}\limits\frac{dt}{t^2}\frac{\sinh{2t}-2t}{\cosh{t}(\alpha\sinh t+t\cosh t)}.
\end{equation}
In the chiral limit $\alpha=-1+a$, where $a=\frac{\pi m^2}{g^2}\to0$ it goes to
\begin{equation}\label{condensate}
    \braket{\bar\psi\psi}\big|_{a\to0}=-\frac{mN_c}{2\pi}\left\{\frac{\pi}{\sqrt{3a}}-\log{\left(\frac{\pi}{a}\right)}+\mathcal{O}(1)\right\}\approx-N_c\sqrt{\frac{g^2}{12\pi}}.
\end{equation}
We expect that in such a limit one massless meson (pion) will arise, that is, the ground-state energy $M_0\to0,$ and hence the even spectral sums $G^{(s)}_+$ blow up (the odd ones $G^{(s)}_-$ remain finite). It is also expected that the Gell-Mann-Oaks-Renner (GMOR) relation \cite{Gell-Mann:1968hlm} linking the pion mass to the quark condensate must be satisfied\footnote{We are grateful to Alexander Gorsky for this remark.}
\begin{equation}\label{GMOR}
    \frac{N_c}{\pi}M_0^2\approx-4m\braket{\bar\psi\psi}.
\end{equation}

To establish the asymptotic form of the spectral sums and extract the ground-state eigenvalue $\lambda_0$ from them, we use the asymptotics of the integrals $\mathtt{u}_{2k-1}(\alpha)$ (see appendix \ref{u-asymptotics-Chiral_limit}). Then the spectral sums admit the expansion in the limit $\alpha\to-1$
\begin{equation}\label{G1-Chiral-limit}
    \begin{aligned}
        &G^{(1)}_+(-1+a)|_{a\to0}=\frac{\sqrt{3}\pi}{\sqrt{a}}+\log{8\pi}-1+\frac{7\zeta(3)}{\pi^2}+\frac{1}{2}(c^{(0)}_1+c^{(0)}_3)+\frac{7\pi}{10}\sqrt{3a}+\mathcal{O}(a),\\
        &G^{(1)}_-(-1+a)|_{a\to0}=\log{8\pi}-3-\frac{7\zeta(3)}{\pi^2}+\frac{1}{2}(c^{(0)}_1-c^{(0)}_3)-\pi\sqrt{3a}+\mathcal{O}(a),\\
    \end{aligned}
\end{equation}
\begin{equation}\label{G2-Chiral-limit}
    \begin{aligned}
        &G^{(2)}_+(-1+a)|_{a\to0}=\frac{3\pi^2}{a}+\frac{\sqrt{3}(56\zeta (3)+\pi^4+4\pi^2(2+c^{(0)}_3))}{2\pi\sqrt{a}}+\mathcal{O}(1),\\
        &G^{(2)}_-(-1+a)|_{a\to0}=\frac{10}{3}+\frac{56\zeta(3)}{3\pi^2}-\frac{124\zeta(5)}{\pi^4}+2(c^{(0)}_3-c^{(0)}_5)-6\pi\sqrt{3a}+\mathcal{O}(a^{3/2}),
    \end{aligned}
\end{equation}
\begin{equation}\label{G3-Chiral-limit}
    \begin{aligned}
        &G^{(3)}_+(-1+a)|_{a\to0}=\frac{3\sqrt{3}\pi^3}{a^{3/2}}+\frac{9(56\zeta(3)+\pi^4+4\pi^2(2+c^{(0)}_3))}{4a}+\frac{\sqrt{3}\pi(21\zeta(3)+2)}{\sqrt{a}}+\frac{6\sqrt{3}\pi c^{(0)}_5}{\sqrt{a}}+
        \\&+\frac{12\sqrt{3}(49\zeta(3)^2+31\zeta(5))}{\pi^3\sqrt{a}}+\frac{28\sqrt{3}(4+3c^{(0)}_3)\zeta(3)}{\pi\sqrt{a}}+\frac{3\sqrt{3}\pi c^{(0)}_3(2+c^{(0)}_3)}{\sqrt{a}}+\frac{\sqrt{3}\pi^3(173+15c^{(0)}_3)}{10\sqrt{a}}+\mathcal{O}(1),\\
        &G^{(3)}_-(-1+a)|_{a\to0}=\frac{4\pi^2}{9}-\frac{104}{15}+\frac{28(15-\frac{46}{\pi^2})\zeta(3)}{45}+\frac{62(16-3\pi^2)\zeta(5)}{3\pi^4}-\frac{1016\zeta(7)}{\pi^6}+(\pi^2-4)c^{(0)}_3-
        \\&-(\pi^2-8)c^{(0)}_5-4c^{(0)}_7-3\sqrt{3a}\pi^3+\mathcal{O}(a^{3/2}).
    \end{aligned}
\end{equation}
From expressions \eqref{G1-Chiral-limit}-\eqref{G3-Chiral-limit}, we indeed see that for $a\to0$, the even spectral sums tend to infinity, while the odd ones remain finite.

Thus, from the spectral sum $G^{(2)}_+$ we can find the leading and subleading asymptotics of the ground state
\begin{equation}
    \lambda_0=\frac{\sqrt{a}}{\sqrt{3}\pi}-\frac{56\zeta(3)+\pi^4+4\pi^2(2+c^{(0)}_3)}{12\pi^4}a+\mathcal{O}(a^{3/2}).
\end{equation}
In principle, using higher sums one can find more terms of the ground-state expansion.

If we look at the asymptotics of the ground state $M_0^2$, we see that it is in agreement with the GMOR formula
\begin{equation} 
    M^2_0\overset{\eqref{M-and-lambda}}{=}2g^2\left\{\frac{\sqrt{a}}{\sqrt{3}}-\frac{56\zeta(3)+\pi^4+4\pi^2(2+c^{(0)}_3)}{12\pi^3}a+\mathcal{O}(a^{3/2})\right\}\approx2g\frac{\sqrt{\pi}m}{\sqrt{3}}\quad\Rightarrow\quad     \frac{N_c}{\pi}M_0^2\approx-4m\braket{\bar\psi\psi}.
\end{equation}
\subsection{The heavy quark limit}
In the heavy quark limit the meson masses $M_n$ approach $2m$ from above. All spectral sums except $G_{\pm}^{(1)}$ are expected to tend to zero at $\alpha\to\infty$, but if we naively look at our formulas, this is not obvious. Using the asymptotics of integrals $\mathtt{u}_{2k-1}(\alpha)$ (see appendix \ref{u-asymptotics-Heavy}), one drives the expansions of the spectral sums at $\alpha\to\infty$
\begin{equation}\label{G1-heavy-limit}
    \begin{aligned}
        &G^{(1)}_+(\alpha)\big|_{\alpha\to\infty}=\log\left(\frac{2\pi^2e^{-\gamma_E-2}}{\alpha}\right)+\frac{\pi^2}{8\alpha}-\frac{\pi^2}{4\alpha^2}+\frac{5\pi^2}{12\alpha^3}-\frac{7\pi^2\left(30+\pi^2\right)}{360\alpha^4}+\frac{\pi^2(15+2\pi^2)}{20\alpha^5}-
        \\&-\frac{11\pi^2(105+35\pi^2+\pi^4)}{1260\alpha^6}+\mathcal{O}\left(\frac{1}{\alpha^7}\right),
        \\
        &G^{(1)}_-(\alpha)\big|_{\alpha\to\infty}=\log\left(\frac{2\pi^2e^{-\gamma_E-2}}{\alpha}\right)-\frac{\pi^2}{8\alpha}+\frac{\pi^2}{12\alpha^2}-\frac{\pi^2}{12\alpha^3}+\frac{\pi^2\left(30+\pi^2\right)}{360\alpha^4}-\frac{\pi^2(15+2\pi^2)}{180\alpha^5}+
        \\&+\frac{\pi^2(105+35\pi^2+\pi^4)}{1260\alpha^6}+\mathcal{O}\left(\frac{1}{\alpha^7}\right),
    \end{aligned}
\end{equation}
\begin{equation}\label{G2-heavy-limit}
    \begin{aligned}
        &G^{(2)}_+(\alpha)\big|_{\alpha\to\infty}=\frac{\pi^2}{3\alpha}+\frac{\pi^4}{16\alpha^2}-\frac{4\pi^4}{15\alpha^3}+\frac{7\pi^4}{9\alpha^4}-\frac{\pi^4(105+4\pi^2)}{63\alpha^5}+\frac{\pi^4(1635+248\pi^2)}{540\alpha^6}+\mathcal{O}\left(\frac{1}{\alpha^7}\right),
        \\
        &G^{(2)}_-(\alpha)\big|_{\alpha\to\infty}=\frac{\pi^2}{3\alpha}-\frac{\pi^4}{16\alpha^2}+\frac{2\pi^4}{15\alpha^3}-\frac{\pi^4}{3\alpha^4}+\frac{2\pi^4(21+\pi^2)}{63\alpha^5}-\frac{\pi^4(21+4\pi^2)}{18\alpha^6}+\mathcal{O}\left(\frac{1}{\alpha^7}\right),
    \end{aligned}
\end{equation}
\begin{equation}\label{G3-heavy-limit}
    \begin{aligned}
        &G^{(3)}_+(\alpha)\big|_{\alpha\to\infty}=\frac{\pi^4}{15\alpha^2}+\frac{\pi^6}{32\alpha^3}-\frac{23\pi^6}{105\alpha^4}+\frac{29\pi^6}{30\alpha^5}+\frac{\pi^6(2296+141\pi^2)}{945\alpha^6}+\mathcal{O}\left(\frac{1}{\alpha^7}\right),
        \\
        &G^{(3)}_-(\alpha)\big|_{\alpha\to\infty}=\frac{\pi^4}{15\alpha^2}-\frac{\pi^6}{32\alpha^3}+\frac{13\pi^6}{105\alpha^4}-\frac{\pi^6}{2\alpha^5}-\frac{\pi^6(665+33\pi^2)}{315\alpha^6}+\mathcal{O}\left(\frac{1}{\alpha^7}\right).
    \end{aligned}
\end{equation}

We see that these asymptotics show that $G^{(s)}_+>G^{(s)}_-$, which is expected from definition of the spectral sums (because $\lambda_{2n}<\lambda_{2n+1}$), but not obvious from \eqref{G1-analytical}-\eqref{G3-analytical}. 
Also, let us emphasize the fact that the lowest energy levels do not give a main contribution to the spectral sums, but a subleading contribution (the main contribution comes from relativistic mesons, whose mass is significantly different from the doubled mass of quarks). This is true since 
\begin{equation}
    \lambda_n\sim M_n^2\sim m^2\sim \alpha\quad\Rightarrow\quad (\lambda_n)^{-s}\sim \alpha^{-s}.
\end{equation}

Note also that the first two leading orders of the spectral sums \eqref{G1-heavy-limit}-\eqref{G3-heavy-limit} coincide with the spectral sums that can be computed using a simplified form of the expressions (3.29), (3.30) from \cite{ZIYATDINOV:2010ModPhysA}\footnote{We are grateful to Alexander Artemev for pointing this out.}, which determine the meson masses in the heavy quark limit 
\begin{equation}\label{Ziyatdinov-heavy-limit}
    \frac{\sinh{2\theta_n}-2\theta_n}{2}=\pi\lambda_{IZ}\left(n-\left[\frac{1}{2}\pm\frac{1}{4}\right]\right)+\mathcal{O}(\lambda^2_{IZ}),\quad n=1,2,3,\dots,
\end{equation}
where $\theta$ is the parameter related with the meson masses 
$\alpha_{IZ}=\frac{M_{IZ}^2}{4m_{IZ}^2}=\cosh^2{\theta}.$ The "$+$" sign corresponds to even energy levels, and "$-$" to odd energy levels. The relation between our notations with those of \cite{ZIYATDINOV:2010ModPhysA}
\begin{equation}
    \alpha_{IZ}=\frac{M_{IZ}^2}{4m^2_{IZ}}=\frac{2\pi^2\lambda\frac{g^2}{\pi}}{4\alpha\frac{g^2}{\pi}}=\frac{\pi^2\lambda}{2\alpha},\quad \lambda_{IZ}=\frac{g^2}{2m_{IZ}^2}=\frac{g^2}{2\alpha\frac{g^2}{\pi}}=\frac{\pi}{2\alpha}.
\end{equation}

The asymptotic expansion of the spectral sums from \eqref{Ziyatdinov-heavy-limit} can be obtained using the Euler-Maclaurin formula
\begin{equation}\label{Euler-Maclaurin}
    \sum_{n=1}^{\infty}f_n\approx\int_0^{\infty}\limits dn f(n)+\frac{1}{2}f(n)\Big|^{\infty}_{0}=\int_{0}^{\infty}\limits d\theta\left|\frac{dn}{d\theta}\right| f(\theta(n))+\int_{\theta_0}^{0}\limits d\theta\left|\frac{dn}{d\theta}\right| f(\theta(n))+\frac{1}{2}f(n)\Big|^{\infty}_{0},
\end{equation}
where
\begin{equation}
    d\theta(\cosh{2\theta}-1)=\frac{\pi^2}{2\alpha}dn+\mathcal{O}\left(\frac{1}{\alpha^2}\right),\quad \theta_0\overset{\text{def}}{=}\theta(0)=-\left(\frac{(6\pm3)\pi^2}{16\alpha}\right)^{\frac{1}{3}}+\frac{(2\pm1)\pi^2}{80\alpha}+\mathcal{O}\left(\frac{1}{\alpha^{5/3}}\right).
\end{equation}
For the first spectral sums $G_{\pm}^{(1)}$, we first rewrite
\begin{equation}
    G^{(1)}_{\pm}(\alpha)=\lim_{n\to\infty}\sum_{k=1}^n\left[\frac{1}{\frac{2\alpha}{\pi^2}\cosh^2{\theta_k}}-\frac{1}{k}\right]=-\gamma_E+\lim_{n\to\infty}\sum_{k=1}^n\left[\frac{1}{\frac{2\alpha}{\pi^2}\cosh^2{\theta_k}}-\frac{\log n}{n}\right].
\end{equation}
Now in the limit $\alpha\to\infty$ we use \eqref{Euler-Maclaurin}
\begin{multline}
    G^{(1)}_{\pm}(\alpha)\big|_{\alpha\to\infty}\approx-\gamma_E+\lim_{n\to\infty}\left\{\int_{\theta_0}^{\theta_n}\limits d\theta\frac{\cosh{2\theta}-1}{\cosh^2{\theta}}+\frac{1}{2}\frac{1}{\frac{2\alpha}{\pi^2}\cosh^2{\theta}}\Biggl|^{\theta_n}_{\theta_0}-\log n\right\}=
    \\=-\gamma_E-\frac{1}{2}\frac{1}{\frac{2\alpha}{\pi^2}\cosh^2{\theta_0}}-2(\theta_0-\tanh{\theta_0})+\lim_{n\to\infty}\left\{2(\theta_n-\tanh{\theta_n})+\frac{1}{2}\frac{1}{\frac{2\alpha}{\pi^2}\cosh^2{\theta_n}}-\log n\right\}.
\end{multline}
After calculations we obtain the desired asymptotics
\begin{equation}
    G^{(1)}_{\pm}(\alpha)\big|_{\alpha\to\infty}\approx \log\left(\frac{2\pi^2e^{-\gamma_E-2}}{\alpha}\right)\pm\frac{\pi^2}{8\alpha}+\mathcal{O}\left(\frac{1}{\alpha^2}\right).
\end{equation}

For the highest spectral sums by means of \eqref{Euler-Maclaurin} we obtain the expressions
\begin{equation}
    G^{(s)}_{\pm}(\alpha)\big|_{\alpha\to\infty}\approx\int_{0}^{\infty}\limits\frac{d\theta}{\frac{\pi^2}{2\alpha}}\frac{\cosh{2\theta}-1}{(\frac{2\alpha}{\pi^2}\cosh^2{\theta})^s}+\int_{\theta_0}^{0}\limits\frac{d\theta}{\frac{\pi^2}{2\alpha}}\frac{\cosh{2\theta}-1}{(\frac{2\alpha}{\pi^2}\cosh^2{\theta})^s}+\frac{1}{2}\frac{1}{(\frac{2\alpha}{\pi^2}\cosh^2{\theta})^s}\Biggl|_{n=0}^{n=\infty},\quad s>1.
\end{equation}
Thus, we have the agreement of this method with our limit formulas \eqref{G1-heavy-limit}-\eqref{G3-heavy-limit}
\begin{equation}
    \resizebox{\textwidth}{!}{$
    \begin{aligned}
        &G^{(2)}_{\pm}(\alpha)\big|_{\alpha\to\infty}\approx\underbrace{\frac{\pi^2}{2\alpha}\int_{0}^{\infty}\limits d\theta\frac{\cosh{2\theta}-1}{\cosh^4{\theta}}}_{\frac{\pi^2}{3\alpha}}-\underbrace{\frac{\pi^2}{2\alpha}\int^{\theta_0}_{0}\limits d\theta\frac{\cosh{2\theta}-1}{\cosh^4{\theta}}}_{\approx -\frac{(2\pm1))\pi^4}{16\alpha^2}}-\underbrace{\frac{\pi^4}{8\alpha^2}\frac{1}{\cosh^4{\theta_0}}}_{\approx\frac{\pi^4}{8\alpha^2}} =\frac{\pi^2}{3\alpha}\pm\frac{\pi^4}{16\alpha^2}+\mathcal{O}\left(\frac{1}{\alpha^3}\right),
        \\
        &G^{(3)}_{\pm}(\alpha)\big|_{\alpha\to\infty}\approx\underbrace{\frac{\pi^4}{4\alpha^2}\int_{0}^{\infty}\limits d\theta\frac{\cosh{2\theta}-1}{\cosh^6{\theta}}}_{\frac{\pi^4}{15\alpha^2}}-\underbrace{\frac{\pi^4}{4\alpha^2}\int_{0}^{\theta_0}\limits d\theta\frac{\cosh{2\theta}-1}{\cosh^6{\theta}}}_{\approx -\frac{(2\pm1)\pi^6}{32\alpha^3}}
        -\underbrace{\frac{\pi^6}{16\alpha^3}\frac{1}{\cosh^6{\theta_0}}}_{\approx\frac{\pi^6}{16\alpha^3}}=\frac{\pi^4}{15\alpha^2}\pm\frac{\pi^6}{32\alpha^3}+\mathcal{O}\left(\frac{1}{\alpha^4}\right).
    \end{aligned}
    $}
\end{equation}
Note that to obtain more summands in the spectral sum expansions from \cite{ZIYATDINOV:2010ModPhysA}, we need to already take into account the correction $\mathcal{O}(\lambda^2_{IZ})$ in \eqref{Ziyatdinov-heavy-limit}, which we ignored, and use more summands in \eqref{Euler-Maclaurin}.
\section{Numerical results}\label{Numerical-results}
\subsection{Method}
To verify our analytical expressions for the spectral sums as well as large $n-$expansion for the spectrum, we use the decomposition of the wave function $\phi(x)$ by the following basis \cite{Hanson:1976ey} 
\begin{equation}\label{Chebyshev_basis}
    \phi(x)=\sum_{n}\phi_n h_n(x) ,\quad h_n(x)=\sin{(n\theta)}=2\sqrt{x(1-x)}U_{n-1}(1-2x), \quad n\geq1,
\end{equation}
where the new variable $\theta\in[0,\pi]$ with $x=\frac{1-\cos{\theta}}{2}$ and $U_n(x)$ are Chebyshev polynomials of the second kind. It allows to reformulate the 't Hooft integral equation into a matrix problem, which in a truncated form (for some sufficiently large $N$) is amenable to machine solution
\begin{equation}
    2\pi^2\lambda\sum_n M_{mn}\phi_n =\sum_n H_{mn}\phi_n, \quad M_{mn}=\braket{h_m|h_n},\quad H_{mn}=\bra{h_m}\mathcal{H}\ket{h_n},
\end{equation}
where the scalar product has a standard form
\begin{equation}
    \braket{h|g}=\int_{0}^{1}\limits dx\; h(x)g(x).
\end{equation}
The matrices $\bm{M}$ and $\bm{H}$ admit the integral form
\begin{equation}
    M_{mn}=\int_0^1\limits dx\;\sin{(m\theta)}\sin{(n\theta)}=
    \begin{cases}
        -\frac{2mn}{m^4+(n^2-1)^2-2m^2(1+n^2)},\quad &m+n=2k,\quad\quad\;\;\, k=1,2,\dots; \\
        0,\quad &m+n=2k-1,\quad k=1,2,\dots
    \end{cases}
\end{equation}
and
\begin{equation}\label{H-matrix-Chebyshev}
    H_{mn}=\underbrace{\int_0^1\limits dx\;\sin{(m\theta)}\sin{(n\theta)}\;\alpha\left(\frac{1}{x}+\frac{1}{1-x}\right)}_{H^{(1)}_{mn}}-\underbrace{\int_0^{1}\limits dx\sin{(m\theta)}\fint_0^1\limits dy\frac{\sin{(n\theta')}}{(x-y)^2}}_{H^{(2)}_{mn},\quad y=\frac{1-\cos{\theta'}}{2}}.
\end{equation}
The first integral $H^{(1)}_{mn}$ in \eqref{H-matrix-Chebyshev} has the form
\begin{equation}
    H^{(1)}_{mn}=2\alpha\int_0^{\pi}\limits d\theta\;\frac{\sin{(m\theta)}\sin{(n\theta)}}{\sin{\theta}}=
    2\alpha\cdot\begin{cases}
        2\sum_{l=\frac{|m-n|}{2}+1}^{\frac{m+n}{2}}\limits\frac{1}{2l-1},\quad &m+n=2k,\quad\quad\;\;\, k=1,2,\dots; \\
        0,\quad &m+n=2k-1,\quad k=1,2,\dots
    \end{cases}
\end{equation}
while the second integral $H^{(2)}_{mn}$ is simply 
\begin{equation}
    H^{(2)}_{mn}=\int_0^1\limits dx\sin{(m\theta)}\fint_0^1\limits dy\frac{\sin{(n\theta')}}{(x-y)^2}=-\pi n\int_0^{\pi}\limits d\theta\;\sin{(m\theta)}\sin{(n\theta)}=-\frac{n\pi^2}{2}\delta_{mn}.
\end{equation}
Thus the matrix elements of $\bm{H}$ are  
\begin{equation}
    H_{mn}=\frac{n\pi^2}{2}\delta_{mn}+H^{(1)}_{mn}.
\end{equation}
This numerical method can also be applied for the case when the quark masses are different $\alpha_1\ne\alpha_2$, details can be found in \cite{Kochergin:2024quv}.
\subsection{Numerical data}
We used $N = 1000$ basis of $h_n$ functions (see \eqref{Chebyshev_basis}), to numerically verify \eqref{WKB-eq}. Tables \ref{Table-WKB-even}, \ref{Table-WKB-odd} show the numerical and analytical values of the first $30$ eigenvalues, as well as the relative error. It can be seen that already at $n=1$ the accuracy is quite high.
\begin{table}[h!]
\begin{center}
    \resizebox{\textwidth}{!}{
    \begin{tabular}{| c | l | c | l|| l | c | l|| l | c | l|| l | c | l|}
    \hline \rule{0mm}{3.6mm}
     $\alpha$ & \multicolumn{3}{c||}{$-0.5$} & \multicolumn{3}{c||}{$-0.1$} & \multicolumn{3}{c||}{$0.1$} & \multicolumn{3}{c|}{$0.5$}\\
    \hline
    $n$ & $\lambda_{n}^{(6)}$ & $\delta\lambda,\%$ &  $\lambda_{n}^{(\text{num})}$ & $\lambda_{n}^{(6)}$ & $\delta\lambda,\%$ &  $\lambda_{n}^{(\text{num})}$ & $\lambda_{n}^{(6)}$ & $\delta\lambda,\%$ &  $\lambda_{n}^{(\text{num})}$ & $\lambda_{n}^{(6)}$ & $\delta\lambda,\%$ & $\lambda_{n}^{(\text{num})}$\\
    \hline
    $0$  & 0.21518 & 3.1 & 0.22186 & 0.23671 & 43.8 & 0.34048 & 0.27578 & 43.6 & 0.39614 & 0.33815 & 48.8 & 0.50322\\
    $2$  & 1.14542 & $3.5\cdot 10^{-3}$ & 1.14538 & 1.33177 & $2.1\cdot10^{-3}$ & 1.33180 & 1.41514 & $2.4\cdot10^{-3}$ & 1.41518 & 1.57042 & $1.6\cdot10^{-3}$ & 1.57044\\
    $4$  & 2.11510 & $7.3\cdot 10^{-6}$ & 2.11511 & 2.32668 & $3.0\cdot10^{-5}$ & 2.32668 & 2.42124 & $3.5\cdot10^{-5}$ & 2.42124 & 2.59703 & $2.9\cdot10^{-6}$ & 2.59703\\
    $6$  & 3.09608 & $9.1\cdot 10^{-5}$ & 3.09608 & 3.32323 & $2.0\cdot10^{-6}$ & 3.32323 & 3.42495 & $2.2\cdot10^{-6}$ & 3.42495 & 3.61415 & $2.2\cdot10^{-6}$ & 3.61415\\
    $8$ & 4.08220 & $9.3\cdot 10^{-5}$ & 4.08221 & 4.32064 & $3.7\cdot10^{-7}$ & 4.32064 & 4.42763 & $2.9\cdot10^{-7}$ & 4.42763 & 4.62684 & $4.7\cdot10^{-7}$ & 4.62684\\
    $10$ & 5.07129 & $8.9\cdot 10^{-5}$ & 5.07130 & 5.31857 & $2.0\cdot10^{-7}$ & 5.31857 & 5.42974 & $8.8\cdot10^{-8}$ & 5.42974 & 5.63694 & $4.6\cdot10^{-8}$ & 5.63694\\
    $12$ & 6.06230 & $8.6\cdot 10^{-5}$ & 6.06231 & 6.31684 & $1.6\cdot10^{-7}$ & 6.31684 & 6.43148 & $5.4\cdot10^{-8}$ & 6.43148 & 6.64535 & $7.4\cdot10^{-8}$ & 6.64535\\
    $14$ & 7.05466 & $8.2\cdot 10^{-5}$ & 7.05466 & 7.31537 & $1.6\cdot10^{-7}$ & 7.31537 & 7.43296 & $4.8\cdot10^{-8}$ & 7.43296 & 7.65254 & $1.2\cdot10^{-7}$ & 7.65254\\
    $16$ & 8.04801 & $8.0\cdot 10^{-5}$ & 8.04802 & 8.31408 & $1.5\cdot10^{-7}$ & 8.31408 & 8.43425 & $4.6\cdot10^{-8}$ & 8.43425 & 8.65883 & $1.3\cdot10^{-7}$ & 8.65883\\
    $18$ & 9.04213 & $7.7\cdot 10^{-5}$ & 9.04214 & 9.31294 & $1.5\cdot10^{-7}$ & 9.31294 & 9.43539 & $4.6\cdot10^{-8}$ & 9.43539 & 9.66442 & $1.4\cdot10^{-7}$ & 9.66442\\
    $20$ & 10.0369 & $7.5\cdot 10^{-5}$ & 10.0369 & 10.3119 & $1.5\cdot10^{-7}$ & 10.3119 & 10.4364 & $4.6\cdot10^{-8}$ & 10.4364 & 10.6695 & $1.5\cdot10^{-7}$ & 10.6695\\
    $22$ & 11.0321 & $7.4\cdot 10^{-5}$ & 11.0321 & 11.3110 & $1.5\cdot10^{-7}$ & 11.3110 & 11.4374 & $4.6\cdot10^{-8}$ & 11.4374 & 11.6740 & $1.5\cdot10^{-7}$ & 11.6740\\
    $24$ & 12.0277 & $7.2\cdot 10^{-5}$ & 12.0277 & 12.3101 & $1.5\cdot10^{-7}$ & 12.3101 & 12.4382 & $4.6\cdot10^{-8}$ & 12.4382 & 12.6782 & $1.5\cdot10^{-7}$ & 12.6782\\
    $26$ & 13.0237 & $7.1\cdot 10^{-5}$ & 13.0237 & 13.3093 & $1.5\cdot10^{-7}$ & 13.3093 & 13.4390 & $4.6\cdot10^{-8}$ & 13.4390 & 13.6821 & $1.5\cdot10^{-7}$ & 13.6821\\
    $28$ & 14.0200 & $6.9\cdot 10^{-5}$ & 14.0200 & 14.3086 & $1.5\cdot10^{-7}$ & 14.3086 & 14.4397 & $4.6\cdot10^{-8}$ & 14.4397 & 14.6857 & $1.6\cdot10^{-7}$ & 14.6857\\
    \hline
    \end{tabular}}
\end{center}
\caption{Values of even eigenvalues of $\lambda_{n}$ from the large-$\lambda$ expansion for several values of $\alpha$. The $\lambda_{n}^{(6)}$ column is obtained from \eqref{WKB-eq} with truncation of the sum behind the term $\propto
\lambda^{-6}$. The ratio $\delta\lambda=\frac{|\lambda_{n}^{(6)}-\lambda_{n}^{(\rm{num})}|}{\lambda_{n}^{(6)}}$ shows the relative error.
The column $\lambda_{n}^{(\text{num})}$ shows the eigenvalues calculated by direct numerical solution of \eqref{'tHooft-eq}.}
\label{Table-WKB-even}
\end{table}
\begin{table}[!h]
\begin{center}
    \resizebox{\textwidth}{!}{
    \begin{tabular}{| c | l | c | l|| l | c | l|| l | c | l|| l | c | l|}
    \hline \rule{0mm}{3.6mm}
     $\alpha$ & \multicolumn{3}{c||}{$-0.5$} & \multicolumn{3}{c||}{$-0.1$} & \multicolumn{3}{c||}{$0.1$} & \multicolumn{3}{c|}{$0.5$}\\
    \hline
    $n$ & $\lambda_{n}^{(6)}$ & $\delta\lambda,\%$ &  $\lambda_{n}^{(\text{num})}$ & $\lambda_{n}^{(6)}$ & $\delta\lambda,\%$ & $\lambda_{n}^{(\text{num})}$ & $\lambda_{n}^{(6)}$ & $\delta\lambda,\%$ &  $\lambda_{n}^{(\text{num})}$ & $\lambda_{n}^{(6)}$ & $\delta\lambda,\%$ &  $\lambda_{n}^{(\text{num})}$\\
    \hline
    $1$  & 0.67348 & $6.5\cdot 10^{-2}$ & 0.67305 & 0.83955 & $6.5\cdot 10^{-2}$ & 0.83901 & 0.91439 & $7.2\cdot10^{-2}$ & 0.91373 & 1.05469 & 0.1 & 1.05359\\
    $3$  & 1.62864 & $4.7\cdot 10^{-5}$ & 1.62864 & 1.82992 & $2.0\cdot 10^{-4}$ & 1.82991 & 1.91985 & $2.4\cdot10^{-4}$ & 1.91984 & 2.08706 & $3.3\cdot10^{-4}$ & 2.08705\\
    $5$  & 2.60498 & $1.0\cdot 10^{-4}$ & 2.60498 & 2.82526 & $7.2\cdot 10^{-6}$ & 2.82526 & 2.92381 & $8.4\cdot10^{-6}$ & 2.92381 & 3.10704 & $1.1\cdot10^{-5}$ & 3.10703\\
    $7$  & 3.58882 & $9.7\cdot 10^{-5}$ & 3.58882 & 3.82211 & $5.4\cdot 10^{-7}$ & 3.82211 & 3.92669 & $7.4\cdot10^{-7}$ & 3.92669 & 4.12131 & $7.8\cdot10^{-7}$ & 4.12131\\
    $9$  & 4.57655 & $9.2\cdot 10^{-5}$ & 4.57656 & 4.81972 & $4.5\cdot 10^{-8}$ & 4.81972 & 4.92894 & $7.8\cdot10^{-8}$ & 4.92894 & 5.13241 & $9.0\cdot10^{-9}$ & 5.13241\\
    $11$ & 5.56667 & $8.8\cdot 10^{-5}$ & 5.56667 & 5.81779 & $1.3\cdot 10^{-7}$ & 5.81779 & 5.93079 & $1.8\cdot10^{-8}$ & 5.93079 & 6.14151 & $1.1\cdot10^{-7}$ & 6.14151\\
    $13$ & 6.55839 & $8.4\cdot 10^{-5}$ & 6.55839 & 6.81617 & $1.5\cdot 10^{-7}$ & 6.81617 & 6.93235 & $3.8\cdot10^{-8}$ & 6.93235 & 7.14921 & $1.3\cdot10^{-7}$ & 7.14921\\
    $15$ & 7.55127 & $8.1\cdot 10^{-5}$ & 7.55127 & 7.81477 & $1.5\cdot 10^{-7}$ & 7.81477 & 7.93371 & $4.3\cdot10^{-8}$ & 7.93371 & 8.15589 & $1.4\cdot10^{-7}$ & 8.15589\\
    $17$ & 8.54502 & $7.9\cdot 10^{-5}$ & 8.54502 & 8.81355 & $1.5\cdot 10^{-7}$ & 8.81355 & 8.93490 & $4.5\cdot10^{-8}$ & 8.93490 & 9.16179 & $1.4\cdot10^{-7}$ & 9.16179\\
    $19$ & 9.53945 & $7.6\cdot 10^{-5}$ & 9.53946 & 9.81246 & $1.5\cdot 10^{-7}$ & 9.81246 & 9.93597 & $4.5\cdot10^{-8}$ & 9.93597 & 10.1671 & $1.5\cdot10^{-7}$ & 10.1671\\
    $21$ & 10.5344 & $7.4\cdot 10^{-5}$ & 10.5344 & 10.8115 & $1.5\cdot 10^{-7}$ & 10.8115 & 10.9369 & $4.6\cdot10^{-8}$ & 10.9369 & 11.1719 & $1.5\cdot10^{-7}$ & 11.1719\\
    $23$ & 11.5299 & $7.3\cdot 10^{-5}$ & 11.5299 & 11.8106 & $1.5\cdot 10^{-7}$ & 11.8106 & 11.9378 & $4.6\cdot10^{-8}$ & 11.9378 & 12.1762 & $1.5\cdot10^{-7}$ & 12.1762\\
    $25$ & 12.5257 & $7.1\cdot 10^{-5}$ & 12.5257 & 12.8097 & $1.5\cdot 10^{-7}$ & 12.8097 & 12.9386 & $4.6\cdot10^{-8}$ & 12.9386 & 13.1802 & $1.5\cdot10^{-7}$ & 13.1802\\
    $27$ & 13.5218 & $7.0\cdot 10^{-5}$ & 13.5218 & 13.8090 & $1.5\cdot 10^{-7}$ & 13.8090 & 13.9394 & $4.6\cdot10^{-8}$ & 13.9394 & 14.1840 & $1.6\cdot10^{-7}$ & 14.1840\\
    $29$ & 14.5182 & $6.9\cdot 10^{-5}$ & 14.5182 & 14.8083 & $1.5\cdot 10^{-7}$ & 14.8083 & 14.9401 & $4.6\cdot10^{-8}$ & 14.9401 & 15.1874 & $1.6\cdot10^{-7}$ & 15.1874\\
    \hline
    \end{tabular}}
\end{center}
\caption{The same for odd eigenvalues.} 
\label{Table-WKB-odd}
\end{table}

To numerically check the spectral sums \eqref{spectral_sums_def}, we use a basis of $N = 1000$ functions $h_n$ (see \eqref{Chebyshev_basis}), which provides an accuracy associated with a cutoff of at least $N^{-s+1}$.  However, this is not a very high accuracy, especially for small values of $s$, since the spectral sums do not converge very quickly. For example, to achieve an accuracy of $5$ decimal places for $G^{(2)}_{\pm}$, we need to sum the first $10^5$ of even/odd spectral values, which requires a basis of $2\cdot10^5$ functions. However, this is not the only difficulty, it turns out that at sufficiently large $n$, the eigenvalues $\lambda^{(\rm{num})}_n$ obtained by this method differ noticeably from the true eigenvalues. When computing \eqref{spectral_sums_def} numerically, starting from a sufficiently large value (through trial and error, we've chosen $n=300$), we use their asymptotic form\footnote{We do not use the highest precision available to us, but only the leading summands with the precision of $\mathcal{O}(\frac{\log n}{n})$. This approximation produces an error of the order of $\mathcal{O}\left(\frac{\log n}{n^{s+2}}\right)$.} \eqref{WKB-eq} instead of the true values of $\lambda_{n}^{(\text{num})}$. Unfortunately, even this trick requires long computations (for each particular $\alpha$ the computation took several tens of minutes), so we calculated eigenvalues up to $10^5$th and used them to numerically compute the spectral sums (see table \ref{G-table}).
\begin{table}[h!]
    \begin{center}
        \resizebox{\textwidth}{!}{
        \begin{tabular}{| c | c | c | c || c | c | c || c | c | c || c | c | c |}
        \hline \rule{0mm}{3.6mm}
        $\alpha$ & \multicolumn{6}{c||}{$-0.5$} & \multicolumn{6}{c|}{$-0.1$}  
        \\
        \hline
        $s$ & $\prescript{\rm{num}}{}G_{+}^{(s)}$ & $\delta G^{(s)}_{+},\%$ & $\prescript{\rm{an}}{}G_{+}^{(s)}$ & $\prescript{\rm{num}}{}G_{-}^{(s)}$ & $\delta G^{(s)}_{-},\%$ & $\prescript{\rm{an}}{}G_{-}^{(s)}$ & $\prescript{\rm{num}}{}G_{+}^{(s)}$ & $\delta G^{(s)}_{+},\%$ & $\prescript{\rm{an}}{}G_{+}^{(s)}$ & $\prescript{\rm{num}}{}G_{-}^{(s)}$ & $\delta G^{(s)}_{-},\%$ & $\prescript{\rm{an}}{}G_{-}^{(s)}$ \\
        \hline
        $1$ & 4.330500 & $3.9\cdot 10^{-2}$ & 4.332200 & 0.786690 & $2.1\cdot10^{-1}$ & 0.788371 & 2.498793 & $6.7\cdot10^{-2}$ & 2.500465 & 0.311769 & $5.3\cdot10^{-1}$ & 0.313428\\
        $2$ & 21.68590 & $4.7\cdot10^{-4}$ & 21.68600 & 3.054819 & $1.1\cdot10^{-3}$ & 3.054853 & 9.725712 & $2.6\cdot10^{-4}$ & 9.725738 & 2.143623 & $1.1\cdot10^{-3}$ & 2.143649\\
        $3$ & 92.42054 & $5.3\cdot10^{-4}$ & 92.42104 & 3.619434 & $4.3\cdot10^{-4}$ & 3.619449 & 25.89887 & $5.9\cdot10^{-7}$ & 25.89887 & 1.945093 & $1.5\cdot10^{-6}$ & 1.945093\\
        $4$ & 413.4268 & $7.1\cdot10^{-4}$ & 413.4297 & 5.047931 & $5.8\cdot10^{-4}$ & 5.047960 & 74.77602 & $7.0\cdot10^{-7}$ & 74.77602 & 2.131654 & $6.8\cdot10^{-7}$ & 2.131654\\
        $5$ & 1861.104 & $8.9\cdot10^{-4}$ & 1861.120 & 7.338725 & $7.3\cdot10^{-4}$ & 7.338779 & 218.8012 & $8.7\cdot10^{-7}$ & 218.8012 & 2.461492 & $8.4\cdot10^{-7}$ & 2.461492\\
        $6$ & 8386.836 & $1.1\cdot10^{-3}$ & 8386.926 & 10.81531 & $8.7\cdot10^{-4}$ & 10.81541 & 642.0568 & $1.0\cdot10^{-6}$ & 642.0568 & 2.895863 & $1.0\cdot10^{-6}$ & 2.895863\\
        \hline
        $\alpha$ & \multicolumn{6}{c||}{$0.1$} & \multicolumn{6}{c|}{$0.5$}  
        \\
        \hline
        $s$ & $\prescript{\rm{num}}{}G_{+}^{(s)}$ & $\delta G^{(s)}_{+},\%$ & $\prescript{\rm{an}}{}G_{+}^{(s)}$ & $\prescript{\rm{num}}{}G_{-}^{(s)}$ & $\delta G^{(s)}_{-},\%$ & $\prescript{\rm{an}}{}G_{-}^{(s)}$ & $\prescript{\rm{num}}{}G_{+}^{(s)}$ & $\delta G^{(s)}_{+},\%$ & $\prescript{\rm{an}}{}G_{+}^{(s)}$ & $\prescript{\rm{num}}{}G_{-}^{(s)}$ & $\delta G^{(s)}_{-},\%$ & $\prescript{\rm{an}}{}G_{-}^{(s)}$ \\
        \hline
        $1$ & 1.984076 & $8.4\cdot10^{-2}$ & 1.985741 & 0.140810 & 1.1 & 0.142462 & 1.275935 & $1.3\cdot10^{-1}$ & 1.277586 & -0.130931 & 1.3 & -0.129294\\
        $2$ & 7.380467 & $3.5\cdot10^{-4}$ & 7.380492 & 1.875509 & $1.4\cdot10^{-3}$ & 1.875535 & 4.819159 & $4.4\cdot10^{-4}$ & 4.819180 & 1.506977 & $1.4\cdot10^{-3}$ & 1.506999\\
        $3$ & 16.56606 & $2.4\cdot10^{-7}$ & 16.56606 & 1.533829 & $1.3\cdot10^{-6}$ & 1.533829 & 8.212447 & $3.4\cdot10^{-7}$ & 8.212447 & 1.035583 & $7.6\cdot10^{-7}$ & 1.035583\\
        $4$ & 40.89774 & $1.7\cdot10^{-7}$ & 40.89774 & 1.529842 & $1.7\cdot10^{-7}$ & 1.529842 & 15.79070 & $3.8\cdot10^{-7}$ & 15.79070 & 0.881717 & $4.1\cdot10^{-7}$ & 0.881717\\
        $5$ & 102.6969 & $2.1\cdot10^{-7}$ & 102.6969 & 1.614770 & $2.1\cdot10^{-7}$ & 1.614770 & 31.10350 & $4.7\cdot10^{-7}$ & 31.10350 & 0.800344 & $4.7\cdot10^{-7}$ & 0.800344\\
        $6$ & 258.8902 & $2.5\cdot10^{-7}$ & 258.8902 & 1.740249 & $2.5\cdot10^{-7}$ & 1.740249 & 61.64925 & $5.6\cdot10^{-7}$ & 61.64925 & 0.744602 & $5.9\cdot10^{-7}$ & 0.744602\\
        \hline
        \end{tabular}}
    \end{center}
    \caption{Numerical and analytical values of the spectral sums \eqref{spectral_sums_def} for several values of $\alpha$. The ratio $\delta G^{(s)}_{\pm}= \left|\frac{\prescript{\rm{num}}{}G_{\pm}^{(s)}-\prescript{\rm{an}}{}G_{\pm}^{(s)}}{\prescript{\rm{an}}{}G_{\pm}^{(s)}}\right|$ shows the relative error.}
    \label{G-table}
\end{table} 

We also did a numerical test of the relation \eqref{dm/dp} (see table \ref{dm/dp-table}). For this purpose, we used the relation \eqref{dprod}, where $\lambda^{(\rm{num})}$ was taken as the first $n$ eigenvalues, and then, starting from $n+1$ and up to $10^4$, we used $\lambda^{(6)}_n$ obtained from the large-$n$ expansion with the accuracy $\mathcal{O}(\frac{\log^6{\mathfrak{n}}}{\mathfrak{n}^6})$.
\begin{table}[h!]
    \begin{center}
        \begin{tabular}{| r | c | c | c | c |}
        \hline \rule{0mm}{3.6mm}
        $\alpha$ & $n$ & $(d_-/d_+)^{\rm{num}}$ & $(d_-/d_+)^{\rm{an}}$ & $\delta d,\%$\\
        \hline
        $-0.5$ & 326 & 0.3182932 & 0.3183099 & $5.2 \cdot10^{-3}$ \\
        $-0.1$ & 325 & 0.4270554 & 0.4270575 & $4.9\cdot10^{-4}$ \\
        $0.1$ & 325 & 0.4721050 & 0.4721299 & $5.3\cdot10^{-3}$\\
        $0.5$ & 324 & 0.5513300 & 0.5513289 & $2.1\cdot10^{-4}$\\
        \hline
        \end{tabular}
    \end{center}
    \caption{Numerical and analytical values for ratio $d_-/d_+$ for several values of $\alpha$. The ratio $\delta d=\left|\frac{(d_-/d_+)^{\rm{num}}-(d_-/d_+)^{\rm{an}}}{(d_-/d_+)^{\rm{an}}}\right|$ shows the relative error.
    \label{dm/dp-table}}
\end{table}
\section{Conclusion}\label{Conclusion}
In this work, we studied the spectrum of meson masses in multicolor QCD$_2$ from the point of view of integrability. We focused on the case of identical quark masses $m_1=m_2$ and showed that the solutions of the 't Hooft's integral equation are closely related to the Baxter's TQ equation. Moreover, we explained how to deal with the technical difficulties arising when trying to construct these solutions in the two limiting cases.  We have found a generalization of the key formula (2.17) in \cite{Fateev:2009jf}, thus generalizing the approach proposed by Fateev, Lukyanov, and Zamolodchikov to the case $\alpha\neq0$. In our work, these are the relations \eqref{2.17-new} and \eqref{2.17-new-2}. In the limiting case of small $\lambda$, we use them to compute the spectral sums $G_{\pm}^{(s)},$ and in the case of large $\lambda$, we obtain the large-$n$ expansion of the spectrum. For the spectral sums we also considered the chiral limit $\alpha\to-1$ and the heavy quark limit $\alpha\to\infty$ and found a complete agreement of our results with other works devoted to the 't Hooft model.

The expressions obtained can be used to find approximate analytic expressions of the lowest-lying energy levels of the bound states. 

In addition, the analytic structure of spectral sums as functions of the complex parameter $\alpha$ is a whole field for theoretical investigations. For example, it is an interesting question to search for critical points of spectral sums and to comprehend the physics behind them. These poles can perhaps be identified with non-unitary CFTs and their identification is an interesting open problem.  

Another separate direction for future research is the Ising model, for which the Bethe-Salpeter equation was derived in the two-particle approximation in \cite{Fonseca:2006au}.

It will be interesting to generalize our results to the case of different quark masses, i.e. for $\alpha_1\neq\alpha_2$. We have obtained preliminary results, and more detailed study of this case will be published elsewhere.
\subsection*{Supplemental material}
We provide Mathematica notebooks \texttt{spectral-sums.nb} and \texttt{Phi.nb} that contain analytic expression for the spectral sums $G^{(s)}_{\pm}$ and coefficients $\Phi^{s}_{\pm}(l)$ in \eqref{quantisation-condition-lambda} for $s\leq7$.  In a separate notebook \texttt{WKB.nb} we collect more orders of the large $n$ formula \eqref{WKB-eq}. In all of these notebooks we use the basic integrals $\mathtt{u}_{2k-1}(\alpha)$, $\mathtt{i}_{2k-1}(\alpha)$ and $\mathtt{i}_{2k}(\alpha)$ given by \eqref{u-def}, \eqref{i2k1-def} and \eqref{i2k-def} respectively.
\subsection*{Acknowledgments}
We acknowledge discussions with Alexander Artemev, Sergei Lukyanov, and Alexander Zamolodchikov. The work of P.M. performed in Landau Institute has been supported by the Russian Science Foundation under the grant 23-12-00333.
\appendix
\section{Check of the relation \eqref{2.17-new}, \eqref{2.17-new-2} and \eqref{Q/Q=D/D}}\label{check-2.17}
Here we provide a non-trivial check of the relations \eqref{2.17-new}, \eqref{2.17-new-2} and \eqref{Q/Q=D/D} utilizing the integral relation \eqref{DD-integral-dev} at $\lambda\rightarrow\infty$. Namely, we compute the differences $F^{(k)}_-(L)-F^{(k)}_+(L)$ and compare them to the ones obtained in section \ref{Large-lambda-expansion}.

In order to compute $F^{(k)}_-(L)-F^{(k)}_+(L)$ using the integral formula \eqref{DD-integral-dev}, it is convenient to represent
\begin{equation}
    Q_{\pm}(\nu|\lambda)=\mathcal{N}_{\pm}(\lambda)\left((-\lambda)^{-\frac{i\nu}{2}}U_{\pm}(\nu|\lambda)\mp
    (-\lambda)^{\frac{i\nu}{2}}U_{\pm}(-\nu|\lambda)\right),
\end{equation}
where
\begin{equation}
    U_{\pm}(\nu|\lambda)=
    T(c^{-1}|\lambda)R_{\pm}(c|\lambda)S_0(\nu)\sum_{k=0}^{\infty}
    \frac{\left(1+\frac{i(\nu+\alpha x)}{2}\right)_k\left(\frac{i(\nu+\alpha x)}{2}\right)_k}{k!}\,(iz)^{-k},
\end{equation}
and $\mathcal{N}_{\pm}(\lambda)$ is the normalization constant absent in \eqref{Q-ansatz}, that has to be chosen to ensure \eqref{Q-normalization-cond}.
Then
\begin{equation}
    \partial_{\lambda}\log\left(\frac{D_+}{D_-}\right)=I_1+I_2,
\end{equation}
where
\begin{multline}
    I_1=\fint_{-\infty}^{\infty}\limits
    \frac{-i\pi\lambda^{-1}}{\nu\coth\frac{\pi\nu}{2}+\frac{2\alpha}{\pi}}
    \frac{(-\lambda)^{i\nu}U_{+}(-\nu)U_{-}(-\nu)-(-\lambda)^{-i\nu}U_{+}(\nu)U_{-}(\nu)}{\sinh\pi\nu}d\nu=
    \\=-2\fint_{-\infty}^{\infty}\limits
    \frac{-i\pi\lambda^{-1}}{\nu\coth\frac{\pi\nu}{2}+\frac{2\alpha}{\pi}}
    \frac{(-\lambda)^{-i\nu}U_{+}(\nu)U_{-}(\nu)}{\sinh\pi\nu}d\nu,
\end{multline}
\begin{equation}
    I_2=\fint_{-\infty}^{\infty}\limits
    \frac{-i\pi\lambda^{-1}}{\nu\coth\frac{\pi\nu}{2}+\frac{2\alpha}{\pi}}
    \frac{U_{+}(-\nu)U_{-}(\nu)-U_{+}(\nu)U_{-}(-\nu)}{\sinh\pi\nu}d\nu.
\end{equation}
The integral $I_1$ can be computed by closing the contour to the lower half plane, where it picks the poles at $\nu=ik$ according to \eqref{S-zeroes-poles}
\begin{multline}\label{I1-values}
    I_1=-2\pi i\left(\frac{1}{2}\textrm{Res}(\dots)\Big|_{\nu=0}+
    \textrm{Res}(\dots)\Big|_{\nu=-i}+\textrm{Res}(\dots)\Big|_{\nu=-2i}+\dots\right)=
    \frac{1}{2\lambda}+\frac{1+2\alpha}{2\pi^2\lambda^2}+\\+
    \frac{4\alpha(1+2\alpha)-4(1+\alpha)L+\alpha(1+\alpha)\mathtt{i}_1(\alpha)}{\pi^4\lambda^3}+\dots
\end{multline}
In order to compute $I_2$, we use the relation \eqref{S0-generic-relation}, which drastically simplifies the integrand. Expanding it as a series of $1/\lambda$ and integrating term-by-term we obtain
\begin{equation}\label{I2-values}
    I_2=\frac{1+\alpha}{2\pi^4\lambda^3}+\frac{(1+\alpha)(3-4L+8\alpha+\alpha\mathtt{i}_1(\alpha))}{4\pi^6\lambda^4}+\dots
\end{equation}
Using \eqref{I1-values} and \eqref{I2-values} we obtain expressions for $F^{(k)}_-(L)-F^{(k)}_+(L)$. The first few of them
\begin{equation}
    F^{(0)}_--F^{(0)}_+=\textrm{const},\quad
    F^{(1)}_--F^{(1)}_+=\frac{1+2\alpha}{2\pi^2},\quad
    F^{(2)}_--F^{(2)}_+=\frac{\alpha(1+\alpha)\mathtt{i}_1(\alpha)+4\alpha(1+2\alpha)-4L(1+\alpha)}{8\pi^4},
\end{equation}
etc, which is in complete agreement with the $F^{(k)}_-(L)-F^{(k)}_+(L)$ obtained in section  \ref{Large-lambda-expansion}. We have verified the difference $F^{(k)}_-(L)-F^{(k)}_+(L)$ up to the order $\lambda^{-7}$ and found a complete agreement as well.
\section{Solution of the shift equation}\label{Shift-equation}
Here we remind a rather general approach to finding solutions to the shift equations, which are of the form
\begin{equation}\label{shift_eq}
    f(\nu+2i)=f(\nu)+h(\nu),
\end{equation}
where $h(\nu)$ is an entire and bounded function of the real variable $\nu$, and we require the solution $f(\nu)$ to be an analytic and bounded function in the strip $\textrm{Im}\,\nu\in[0,2]$. Such requirements for the solution guarantee its uniqueness up to a shift. It can be found as follows. Consider the following integral
\begin{equation}
    I(\nu)=\frac{1}{4i}\oint_{\mathcal{C}}\limits f(t)\left(\tanh\frac{\pi(t-\nu)}{2}-
    \tanh\frac{\pi t}{2}\right)dt,
\end{equation}
where the contour $\mathcal{C}$ is taken as  
\begin{equation*}
    \psfrag{x}{$\nu+i$}
    \psfrag{0}{$0$}
    \psfrag{1}{$i$}
    \psfrag{2}{$2i$}
    \includegraphics[width=0.9\textwidth]{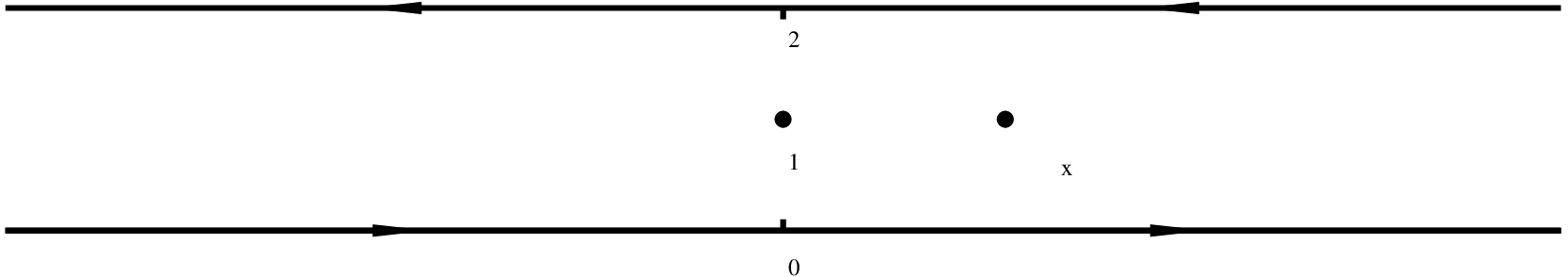}
\end{equation*}
On one hand by Cauchy theorem one has (we use the fact that $f(\nu)$ is an entire function in the strip)
\begin{equation}
    I(\nu)=f(\nu+i)-f(i)
\end{equation}
and on the other rewriting the integral as $\int_{-\infty}^{\infty}\limits-\int_{-\infty+2i}^{\infty+2i}\limits$ and using \eqref{S0-equation} one finds
\begin{equation}
    I(\nu)=\frac{i}{4}\int_{-\infty}^{\infty}\limits h(t)
    \left(\tanh\frac{\pi(t-\nu)}{2}-
    \tanh\frac{\pi t}{2}\right)dt,
\end{equation}
and hence
\begin{equation}
    f(\nu+i)=f(i)+\frac{i}{4}\int_{-\infty}^{\infty}\limits h(t)
    \left(\tanh\frac{\pi(t-\nu)}{2}-
    \tanh\frac{\pi t}{2}\right)dt.
\end{equation}
The described method also allows us to solve multiplicative equations, which are reduced to additive equations by logarithmization. 
\begin{equation}
    F(\nu+2i)=H(\nu)F(\nu)\quad\Rightarrow\quad \log F(\nu+2i)=\log F(\nu)+\log H(\nu).
\end{equation}
Only now, the desired solutions $F(\nu)$, in addition to the properties previously required, must have an additional one, namely, to have no zeros in the strip $\textrm{Im}\,\nu\in[0,2]$.

We note that the functional relation \eqref{Psi-equation} obeyed by the function $\uppsi_{\alpha}(\nu)$ formally violates the conditions specified above. That is, the shift function $h(\nu)$ has a simple pole at the origin $\nu=0$. In order to preserve the analytic properties, one has to shift the integration contour by $-i\epsilon$.     
\section{Asymptotics of the integrals \eqref{u-def}}\label{u-asymptotics}
Here, we describe the asymptotic behavior of the integrals $\mathtt{u}_{2k-1}(\alpha)$ defined by \eqref{u-def}.
\subsection{Chiral limit}\label{u-asymptotics-Chiral_limit}
In the chiral limit we represent $\alpha=-1+a \to-1$, where $a=\frac{\pi m^2}{g^2}\to0.$

As $a\to 0$, the poles $\pm t_1^*=\pm\frac{i\pi}{2}\nu^*_1$ of the integrand  of $\mathtt{u}_{2k-1}(\alpha)$ lying in the strip $\textrm{Im\,}\nu\in[-2,2]$, that is the roots of the transcendental equation \eqref{main-trancendental-equation}, admit the expansion
\begin{equation}
    \quad t^*_1=i\sqrt{3a}\left[1-\frac{1}{10}a+\frac{9}{1400}a^2+\frac{9}{14000}a^3-\frac{1467}{8624000}a^4-\frac{6399}{800800000}a^5+\mathcal{O}\left(a^6\right)\right]\to0.
\end{equation}
Since these poles approach the real axis at $a\to 0$, we deform the integration contour and write (assuming that the contour passes above the pole $t_1^*$)
\begin{equation}
    \mathtt{u}_{2k-1}(\alpha)\big|_{\alpha\to-1}=\underbrace{2\pi i\, \underset{t=t^*_1}{\text{Res\;\;}}\frac{\sinh^2t}{t\cosh^{2k-1}t\big(\alpha\sinh t+t\cosh t\big)}}_{\mathtt{u}_{2k-1}^{(1)}(\alpha)\big|_{\alpha\to-1}}+\underbrace{\int_{-\infty+i\epsilon}^{\infty+i\epsilon}\limits\frac{\sinh^2t}{t\cosh^{2k-1} t \big(\alpha\sinh t+t\cosh t\big)}dt}_{\mathtt{u}_{2k-1}^{(2)}(\alpha)\big|_{\alpha\to-1}}.
\end{equation}
The first summand is a series over half-integer powers of the parameter $a$ 
\begin{multline}
    \mathtt{u}_{2k-1}^{(1)}(-1+a)\big|_{a\to 0}=\frac{2\pi i\sinh^2t}{t\cosh^{2k-1}t\big(a\cosh t+t\sinh t\big)}\Biggl|_{t=t^*_1}=
    \\=\frac{\sqrt{3}\pi}{\sqrt{a}}\left[1+\frac{1}{10}(30k-23)a+\frac{1}{56}\left(252k^2-336k+125\right)a^{2}+\mathcal{O}\left(a^{3}\right)\right],
\end{multline}
and the second one is a series over integer powers
\begin{equation}
    \mathtt{u}_{2k-1}^{(2)}(-1+a)\big|_{a\to 0}=\sum_{l=0}^\infty c^{(l)}_{2k-1} a^l,\quad c^{(l)}_{2k-1}\overset{\textrm{def}}{=}(-1)^l\int_{-\infty+i\epsilon}^{\infty+i\epsilon}\limits\frac{\sinh^{l+2}t}{t\cosh^{2k-1} t \big(t\cosh t-\sinh t\big)^{l+1}}dt,
\end{equation}
where coefficients $c^{(l)}_{2k-1}$ may be calculated numerically. 

Thus, we have the asymptotic of $u_{2k-1}(\alpha)$ in the chiral limit
\begin{equation}
    \mathtt{u}_{2k-1}(-1+a)\big|_{a\to0}=\frac{\sqrt{3}\pi}{\sqrt{a}}+c^{(0)}_{2k-1}+\frac{\sqrt{3}\pi(30k-23)}{10}\sqrt{a}+c^{(1)}_{2k-1}a+\mathcal{O}\left(a^{\frac{3}{2}}\right).
\end{equation}
\subsection{Heavy quark mass limit}\label{u-asymptotics-Heavy}
By the heavy quark mass limit we mean the case where $\alpha\to\infty$. In this limit for $k>1$ we can write the expansion by inverse powers of $\alpha$ (for $k=1$ the first summand will diverge)
\begin{equation}
    \mathtt{u}_{2k-1}(\alpha)\big|_{\alpha\to\infty}=\frac{1}{\alpha}\int_{-\infty}^{\infty}\limits dt\frac{\sinh{t}}{t\cosh^{2k-1}{t}}\sum_{l=0}^{\infty}\left(-\frac{t\coth{t}}{\alpha}\right)^l.
\end{equation}
The integrals arising in the summands are calculated analytically. For example
\begin{equation}
    \begin{aligned}
        &\mathtt{u}_3(\alpha)=\frac{14\zeta(3)}{\pi^2\alpha}-\frac{2}{\alpha^2}+\frac{\pi^2}{4\alpha^3}-\frac{\pi^2}{3\alpha^4}+\frac{\pi^2}{2\alpha^5}-\frac{\pi^2(30+\pi^2)}{45\alpha^6}+\frac{\pi^2(15+2\pi^2)}{18\alpha^7}
        +\mathcal{O}\left(\frac{1}{\alpha^8}\right),\\
        &\mathtt{u}_5(\alpha)=\frac{14\pi^2\zeta(3)+186\zeta(5)}{3\pi^4\alpha}-\frac{4}{3\alpha^2}+\frac{\pi^2-4}{4\alpha^3}-\frac{\pi^2}{6\alpha^4}+\frac{\pi^2(16-\pi^2)}{32\alpha^5}-\frac{2\pi^2(15-\pi^2)}{45\alpha^6}+\frac{\pi^2(15-\pi^2)}{18\alpha^7}+
        \\&+\mathcal{O}\left(\frac{1}{\alpha^8}\right),\\
        &\mathtt{u}_7(\alpha)=\frac{2(56\pi^4\zeta(3)+930\pi^2\zeta(5)+5715\zeta(7))}{45\pi^6\alpha}-\frac{16}{15\alpha^2}+\frac{3\pi^2-16}{12\alpha^3}-\frac{12+\pi^2}{18\alpha^4}+\frac{\pi^2(12-\pi^2)}{16\alpha^5}-
        \\&-\frac{\pi^2(240-19\pi^2)}{360\alpha^6}+\frac{\pi^2(480-128\pi^2+9\pi^4)}{576\alpha^7}+\mathcal{O}\left(\frac{1}{\alpha^8}\right).
    \end{aligned}
\end{equation}

For $k=1$ we use
\begin{equation}
    \mathtt{u}_{1}(\alpha)=2\int_{0}^{\infty}\limits\frac{dt}{t}\frac{\theta(t-1)}{t+\alpha}+2\int_{0}^{\infty}\limits\frac{dt}{t}\left\{\frac{\tanh^2{t}}{t+\alpha\tanh{t}}-\frac{\theta(t-1)}{t+\alpha}\right\},
\end{equation}
where $\theta$ is the Heaviside Step Function. The first term is trivial and equals $\frac{2\log{(1+\alpha)}}{\alpha}$. The second term admits regular expansion at $\alpha\to\infty$
\begin{equation}
    \int_{0}^{\infty}\limits\frac{dt}{t}\left\{\frac{\tanh^2{t}}{t+\alpha\tanh{t}}-\frac{\theta(t-1)}{t+\alpha}\right\}\Biggl|_{\alpha\to\infty}=\frac{1}{\alpha}\sum_{l=0}^{\infty}\frac{(-1)^l}{\alpha^l}\int_{0}^{\infty}\limits dt\left\{\left(\frac{t}{\tanh{t}}\right)^{l-1}-t^l\theta(t-1)\right\}.
\end{equation}
As a result, we have the following asymptotic form for $u_1(\alpha)$
\begin{equation}
    \mathtt{u}_{1}(\alpha)\big|_{\alpha\to\infty}=\frac{2\log{\alpha}}{\alpha}+\frac{2\log{(\frac{4e^{\gamma_E}}{\pi})}}{\alpha}+\frac{\pi^2}{6\alpha^3}-\frac{\pi^2}{3\alpha^4}+\frac{\pi^2(30+\pi^2)}{60\alpha^5}-\frac{2\pi^2(15+2\pi^2)}{45\alpha^6}
    +\mathcal{O}\left(\frac{1}{\alpha^8}\right).
\end{equation}
\bibliographystyle{MyStyle}
\bibliography{MyBib}

\end{document}